\documentclass{SciPost}

\hypersetup{
    colorlinks,
    linkcolor={red!50!black},
    citecolor={blue!50!black},
    urlcolor={blue!80!black}
}

\usepackage[bitstream-charter]{mathdesign}
\usepackage{eso-pic}
\usepackage{mdframed}
\usepackage{slashed}

\usepackage[utf8]{inputenc}
\usepackage{textcomp}
\DeclareSymbolFont{usualmathcal}{OMS}{cmsy}{m}{n}
\DeclareSymbolFontAlphabet{\mathcal}{usualmathcal}

\fancypagestyle{SPstyle}{
\fancyhf{}
\fancyfoot[C]{\textbf{\thepage}}
}
\newcommand{\be}{\begin{equation}}
\newcommand{\ee}{\end{equation}}
\newcommand{\la}{\label}
\newcommand{\eq}[1]  {(\ref{#1})  }
\newcommand{\IM}{\mbox{Im}\,}
\newcommand{\RE}{\mbox{Re}\,}
\newcommand{\p}{\partial}
\newcommand{\vv}{{\rm v}}

\usepackage{xcolor}
\newcommand{\VM}[1]{#1}

\begin{document}

\pagestyle{SPstyle}

\begin{center}{\Large \textbf{\color{scipostdeepblue}{
Finite-gap potentials as a semiclassical limit of the thermodynamic  Bethe Ansatz\\
}}}\end{center}

\begin{center}\textbf{
Valdemar Melin\textsuperscript{1},
Paul Wiegmann\textsuperscript{2} and
Konstantin Zarembo\textsuperscript{3,4}
}\end{center}

\begin{center}
{\bf 1} Department of Physics, KTH Royal Institute of Technology, Stockholm, Sweden\\
{\bf 2} Leinweber  Institute for Theoretical Physics, University of Chicago, Chicago, USA\\
{\bf 3} Nordita, KTH Royal Institute of Technology and Stockholm University, Stockholm, Sweden\\
{\bf 4} Niels Bohr Institute, Copenhagen University, Copenhagen, Denmark
\end{center}

\section*{\color{scipostdeepblue}{Abstract}}
\textbf{\boldmath{%
We show that the semiclassical limit of thermodynamic Bethe Ansatz equations naturally reconstructs the algebro-geometric spectra of finite-gap periodic potentials. This correspondence is illustrated using the traveling-wave (snoidal) solution of the defocusing modified Korteweg--de Vries equation. In this framework, the Bethe-root distribution of the associated quantum field theory yields an Abelian differential of the second kind on the elliptic Riemann surface specified by the spectral endpoints, a structure central to the algebro-geometric theory of solitons. The semiclassical parameter is identified with the large-rank limit of the internal symmetry group $O(2N)$ of the underlying quantum field theory (the Gross-Neveu model with a chemical potential). Our analysis indicates that the analytic structure of the spectrum is dictated solely by the Dynkin diagram $D_N$ and its large-rank limit ($D_\infty$), independently of the particular integrable model used to realize it.
}}
\begin{flushright}
    \emph{Dedicated to the memory of Igor Krichever}
\end{flushright}

\vspace{\baselineskip}






\section{Introduction and Background}
\subsection{Finite-gap potentials and the Peierls phenomenon}
The spectrum of a one-dimensional
\VM{Schr\"odinger operator}
\be
 -\frac{d^2}{dx^2} + u(x) \,
\la{1}
\ee
with a general periodic potential \( u(x) \), consists of an infinite sequence of spectral bands. The bands are admissible energy intervals separated by an infinite sequence of gaps. 
There exists, however, a distinguished class of potentials that gives rise to only a finite number of gaps and, correspondingly, a finite number of spectral bands.

It has been known for nearly fifty years that isospectral deformations of finite-gap potentials are periodic solutions of nonlinear integrable equations \cite{Novikov}. For example, the Korteweg--de Vries (KdV) equation, a paradigmatic example in soliton theory,
\be
u_t - 6 u u_x + u_{xxx} = 0\,
\la{KdV}
\ee
describes a one-parameter family of isospectral deformations of a finite-gap potential of the Schr\"odinger operator \eq{1}. 

The concept of finite-gap potentials is central to the algebro-geometric approach to soliton theory  \cite{list}. At the same time, the role of finite-gap solutions in the quantum counterpart of soliton theory, namely quantum integrable models and their algebro-geometric description, is yet to be understood. This paper aims to shed some light on this issue. We will show how a finite-gap solution emerges as the semiclassical limit of the thermodynamic Bethe Ansatz. 

Our approach relies on an important property of finite-gap potentials: they minimize the energy functional of certain electronic systems modeling the Peierls instability, a key phenomenon in condensed matter physics. 
Formulated in 1930 and published in 1954 \cite{Peierls}, Rudolf Peierls showed that, in one spatial dimension, electrons induce a periodic displacement of crystalline ions with a period commensurate with the mean distance between electrons. 
In turn, the ion displacement modifies the electronic spectrum. The spectrum near the Fermi energy is no longer continuous, instead exhibiting a sequence of gaps separated by fully occupied or empty conducting bands. 
In the early 1980s, it was shown that the Peierls periodic displacements correspond to finite-gap potentials 
\cite{Belokolos,Horovitz1981,Bra_G_K1980}.
The finite-gap potential $\Delta(x)$ minimizes the energy of the electron-phonon system:  
\begin{align}
 \mathcal{E}[\Delta] =\frac{1}{2\lambda} \int \Delta^2 dx  +\sum_{ E<\mu}E \,. \la{5}
\end{align}
Here, $\Delta$ represents the ion displacement, the first term is the elastic energy with \(\lambda>0\), and the second term is the electronic energy, where the sum runs over energy levels consecutively occupied by $N_s$ electrons. The sum is bounded by the chemical potential $\mu_s$, a function of $N_s$. The two terms, having opposite signs, compete, forcing $\Delta$ to be periodic with period $L/N_s$, where $L$ is the system length. Introducing the density of states $\rho(E) dE$, normalized by $N_s=\int \rho(E)dE$, we replace the second term by the integral $\int E \rho(E)dE$.

We focus on a class of finite-gap potentials corresponding to the Peierls phenomenon near commensurate half-filling. In this case, the energy levels in \eq{5} are the eigenvalues of the one-dimensional Dirac operator:
\begin{align}
 H_D \psi_{E} = E \psi_{E},\quad  H_D = -i\alpha \partial_x + \beta \Delta  \,, \la{6}
\end{align}
where \( \alpha, \beta \) are Dirac matrices satisfying \( \alpha^2 = \beta^2 = 1\) and \(\beta \alpha=-\alpha \beta\). The minimum of the energy \eq{5} is achieved under the self-consistency relation 
\begin{align}
 \Delta(x) =\lambda \sum_{ E<\mu} \psi_{E}^\ast(x)\beta\psi_{E}(x) \,. \la{61}
\end{align}

In essence, the finite-gap state is completely determined by the spectrum of the Hamiltonian, which in turn is determined by the density of states $\rho(E)$. This is the primary object of our study.

It is instructive to formulate the problem as a quantum field theory. The Lagrangian reads 
\begin{align}
 \mathcal{L} = \bar \psi(i\slashed{D}-\Delta)\psi  - \frac{1}{2\lambda}\Delta^2 + \frac{M}{2}\dot\Delta^2\,, \la{3}
\end{align}
where the fermionic field $\psi$ and  $\bar\psi:=\psi^\dag\gamma^0$, and the gap-field \( \Delta(x)\) are both quantum fields. Here, $\slashed{D}=\gamma^0(\p_0-i\mu)+\gamma^1\p_x$, $\gamma^0=\beta, \ \gamma_1=\beta\alpha$ are Dirac gamma matrices, and the chemical potential $\mu$ is counted  from zero energy. The ion mass \( M \) (in units of the electron mass) is assumed to be large. In this limit $M\to\infty$, the dynamics of \( \Delta \) is adiabatic, obeying condition\eq{61}. Hence, the problem is reduced to the Peierls model (\ref{5},\ref{6},\ref{61}).

Isospectral deformations of the finite-gap potential $\Delta(x)$ obey integrable hierarchies. In our case, it is the hierarchy generated by the \textit{defocusing} modified KdV (mKdV) equation (note the sign of the nonlinear term):
\begin{align}
 \Delta_t - 6\Delta^2 \Delta_x + \Delta_{xxx} = 0\,, \la{mKdV}
\end{align}
and the Dirac operator, the Hamiltonian $H_D$ of the Peierls model, appears as the Lax (Zakharov-Shabat) operator of mKdV \cite{Ablowitz,Zakharov}. 

\begin{figure}[htbp] 
   \centering
   \includegraphics[width=2in]{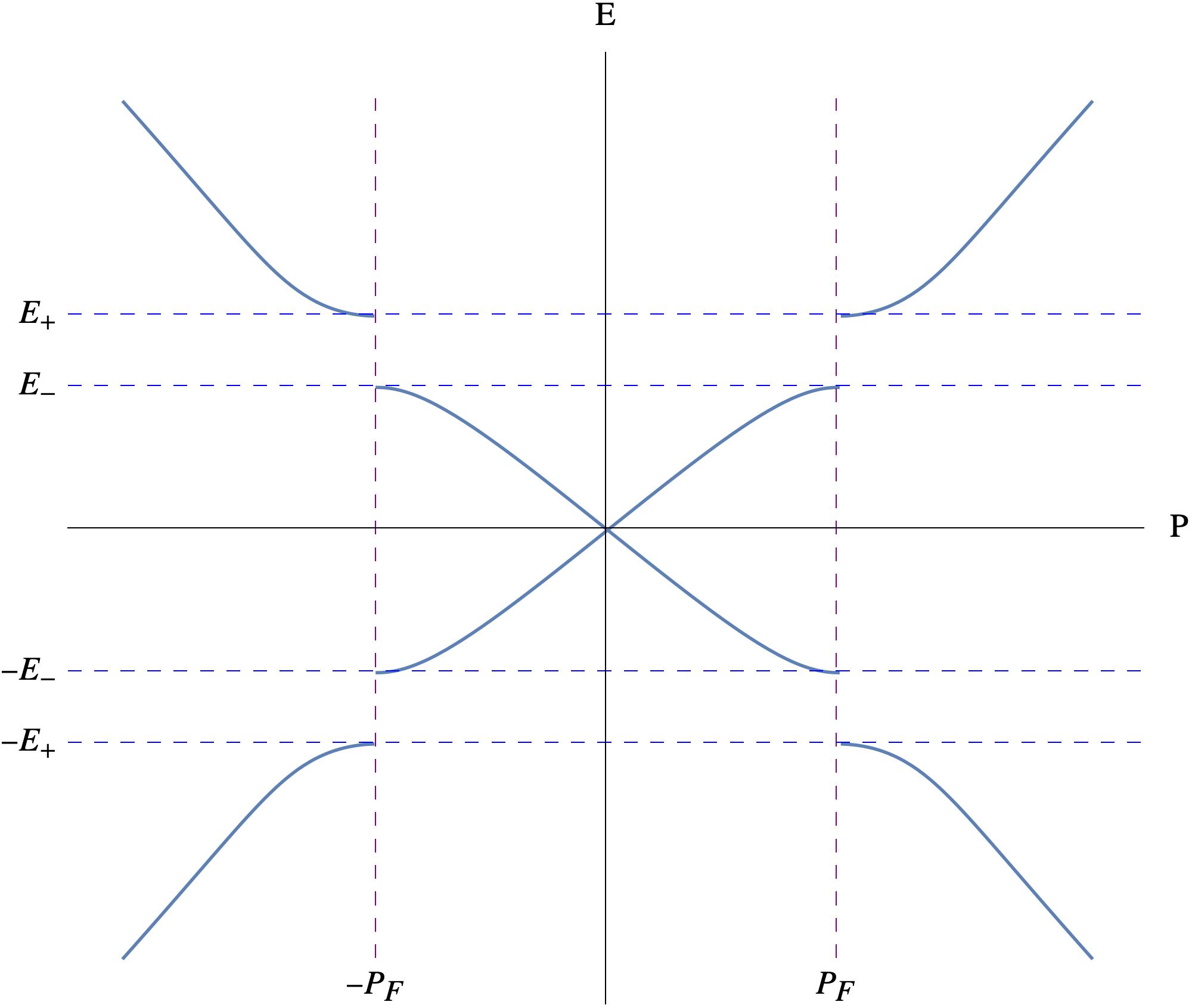} 
   \caption{The spectrum of the traveling (snoidal) wave: the central band $(-E_-,E_-)$  is formed by hybridized zero modes of Dirac operator localized on kinks;  the most upper/lower bands   $(-\infty,-E_+),\  (E_+,\infty)$   correspond to elementary fermions; the spectrum ends on the Fermi momentum $\pm P_F$. The intervals $(-E_+, -E_-),\ (E_-, E_+)$ are gaps.}
   \label{spectrum}
\end{figure}

If the adiabatic condition is removed while retaining integrability, one may be able to identify the quantum version of finite-gap solutions. This is the approach taken in this paper. We identify (i) an integrable quantum field theory whose adiabatic limit yields (\ref{6},\ref{61}), (ii) the quantum state corresponding to a finite-gap potential, and then (iii) compute the spectrum density. We demonstrate this approach for the simplest finite-gap potential, the traveling wave of mKdV, known as the {\it snoidal} wave. 
The spectrum of the snoidal wave consists of two symmetrical gaps, as shown in Fig.~\ref{spectrum}. No immediate obstacles appear for extending this approach to a more general finite-gap setting.

If the adiabatic condition is relaxed by assuming a finite \(M\) in \eq{3}, the problem is not known to be integrable. However, a natural modification retains integrability. We just assume that a fermion state is an $N$-component multiplet $\psi_a:\ a=1,\dots, N$. Then the limit $N\to \infty$ restores the adiabatic condition. The Lagrangian of this integrable model, known as the \(N\)-species Gross-Neveu (GN) model \cite{GN}, reads
\begin{align}
 \mathcal{L}=\sum_{a=1}^N\overline{\psi}_ai\slashed{D}\psi_a+\tfrac {\lambda}{2}\left(\sum_{a=1}^N\overline{\psi}_a\psi_a\right)^2\,.\la{16}
\end{align}
It could be brought closer to the Peierls model by expressing the interaction  via a mediating  gap-field :
\begin{align}
 \mathcal{L}=\sum_{a=1}^N\overline{\psi}_a(i\slashed{D}-\Delta)\psi_a-\tfrac 1{2\lambda}\Delta^2 \,.\la{17}
\end{align}

The model reduces to the Peierls model \eq{3} in the limit of a large number of fermionic species \(N \to \infty\), effectively imposing an adiabatic condition \eq{61}. We see it by integrating out the auxiliary \(N-1\) fermionic species. This  replaces the term \(M\dot\Delta^2\) in \eq{3} with \(N\log {\rm Det}(i\slashed{D}+\Delta)\). Comparing with the gradient expansion of the determinant, one identifies $M$ with $N/\mu^2$. 

Unlike the Peierls model \eq{3}, the GN model is integrable \cite{ZZ,Karowski}. Analyzing the finite-\(N\) solution, we find that the large-\(N\) limit reproduces the finite-gap spectrum of the Dirac operator. 

The GN model is an early example of an integrable system invariant under a Lie group, the \(G\)-invariant integrable systems, with \(G=O(2N)\). This becomes evident when representing each Dirac spinor in \eq{16} as a pair of Majorana spinors $\left(\RE\psi_a, \IM\psi_a\right)$ .

A defining feature of \(G\)-systems is that their complete quantum solution
can be read directly from the Dynkin scheme---\(D_N\) in the present case---which
encodes the fundamental representations of \(G\) and their automorphisms.
In particular, the particle content of the \(D_N\) model is in direct
correspondence with the fundamental representations of \(D_N\), including
the vector representation, antisymmetric tensor representations of ranks
\(r=2,\dots,N-2\), and the two chiral spinor representations.

 A semiclassical limit depends on the chosen quantum state. The state we consider is a thermodynamic state that consists of a large number $N_s$ (increased proportionally to the system size $L$), of pairs of spinors with opposite chirality. We show that the large $N$ limit of this state corresponds to the traveling wave and that the lower-level excitations are only vector particles.    They are {\it minuscule} representations of $D_N$, and they retain the semiclassical meaning.
\begin{figure}[htbp] 
   \centering
   \includegraphics[width=2in]{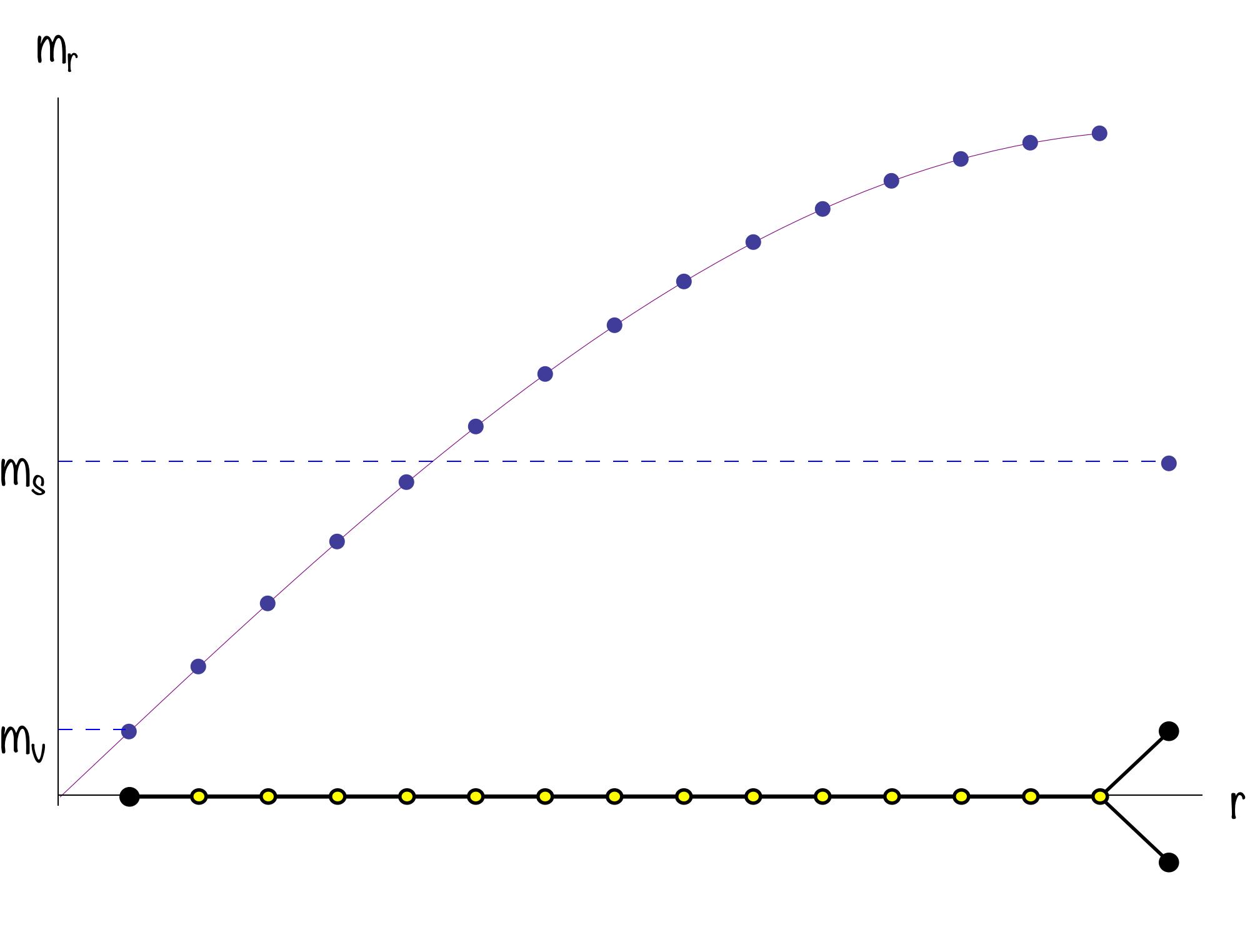} 
   \caption{The mass spectrum of the Dynkin scheme $D_N$: $m_\vv$ and $m_s$ are the masses of multiplets of minuscule representations: the vector and spinors with opposite chirality--the marked nodes on the scheme.}
   \label{mass_spectrum}
\end{figure}

In the semiclassical limit, the vector particles, referred to as elementary
fermions, represent the eigenstates of the Dirac operator~\eq{3}, while the
spinors correspond to half-soliton--like bright or dark kinks, depending on
their chirality~\cite{Dashen}. In the quantum theory at finite
\(N\), vector particles appear as bound states of spinors, and a spinor itself
can be viewed as a bound state of a spinor and a vector particle. This pattern
changes in the semiclassical limit. As \(N\) increases, the mass of a spinor
becomes \(N\) times the mass of a vector particle, as depicted in
Fig.~\ref{snoidal}; consequently, kinks can no longer be treated as particle-like
excitations (see Sec.~\ref{6.1}). Nevertheless, elementary fermions can be
scattered by a kink and may become localized on it. 

There are other states that lead to the same limit. They are large rank-$r$ antisymmetric tensors:  $r/N\to 1$. They could be considered as a pair of dark and bright kinks, Fig.~\ref{snoidal}.


\VM{Like in the case of the spinor-built ground state, the case we primarily focus on, only particles with $r/N\to 0$ or $r/N\to 1$ survive the semiclassical limit. High-rank antisymmetric states in the middle of the Dynkin diagram, where both  $r$ and $N-r$ are of order $N$, disappear from the spectrum because they are too heavy. A lower rank-$r$ tensor decouples into $r$ elementary fermions, while an excitation corresponding to a tensor of rank $N-\bar{r}$ decouples into $\bar{r}$ elementary "holes". }

In the next sections, we briefly summarize the results of (i) the algebro-geometric arguments yielding the symmetric 2-gap spectrum, Sec.~\ref{1.2.2} and  (ii) the scattering problem of Dirac fermions by a kink, Sec.~\ref{PT}. These results serve as benchmarks for comparison with the large $N$ limit of the exact solution of the $O(2N)$-invariant quantum problem.

\subsection{2-gap potential: mKdV traveling  wave\la{1.2}}
 

When the chemical potential in \eq{5} is zero, the adiabatic condition \eq{61} implies that the optimal gap function is uniform, and is given by  
\begin{align} 
\Delta_0= \Lambda e^{-\pi/(N-1)\lambda}\,. \la{811}
\end{align}
Its value is determined by the UV cutoff \( \Lambda \) (the Fermi energy of the underlying lattice model). 
$\Delta_0$ sets the overall scale of momentum and energy. Consequently, 
the spectrum of the Dirac equation \( P^2 = E^2 - \Delta_0^2 \) contains a single symmetric gap \( (-\Delta_0, \Delta_0) \).

At a non-zero chemical potential, the Peierls phenomenon induces a periodic gap function. The simplest example is a traveling wave $\Delta(x+ct)$ of mKdV \eq{mKdV}, which satisfies the ODE \cite{FootnoteC}: 
\begin{align}
 c \Delta_x - 6\Delta^2 \Delta_x + \Delta_{xxx} = 0 \,. \la{81}
\end{align}
The periodic solution of this equation is the \textit{snoidal} wave \cite{Ablowitz}:
\begin{align}
 \Delta(x, k) = \Delta_0 k^{1/2} \, \text{sn}( k^{-1/2} \xi, k),\quad \xi=\Delta_0 x\,, \la{7}
\end{align}
with elliptic modulus $k$ determined by the velocity in \eq{81} as \( c = \Delta_0^2\,(
k^{-1} + k) \). \footnote{\VM{A scaling transformation $\Delta(x)\to \lambda\Delta_0 (\lambda x)$ leaves solutions of \eq{81} invariant with an arbitrary $\lambda$. The condition \eq{61} fixes $\lambda=\Delta_0$.}}
The period of the wave matches the mean distance between fermions. Equating it to the period of the elliptic sine 
\begin{align} (L /{N_s})\Delta_0=4 k^{1/2} K(k)\, \la{131}
\end{align}
 where $K(k)$ is the complete elliptic integral of the first kind, determines the elliptic modulus $k$ (albeit implicitly).  

\begin{figure}[htbp] 
   \centering
   \includegraphics[width=2in]{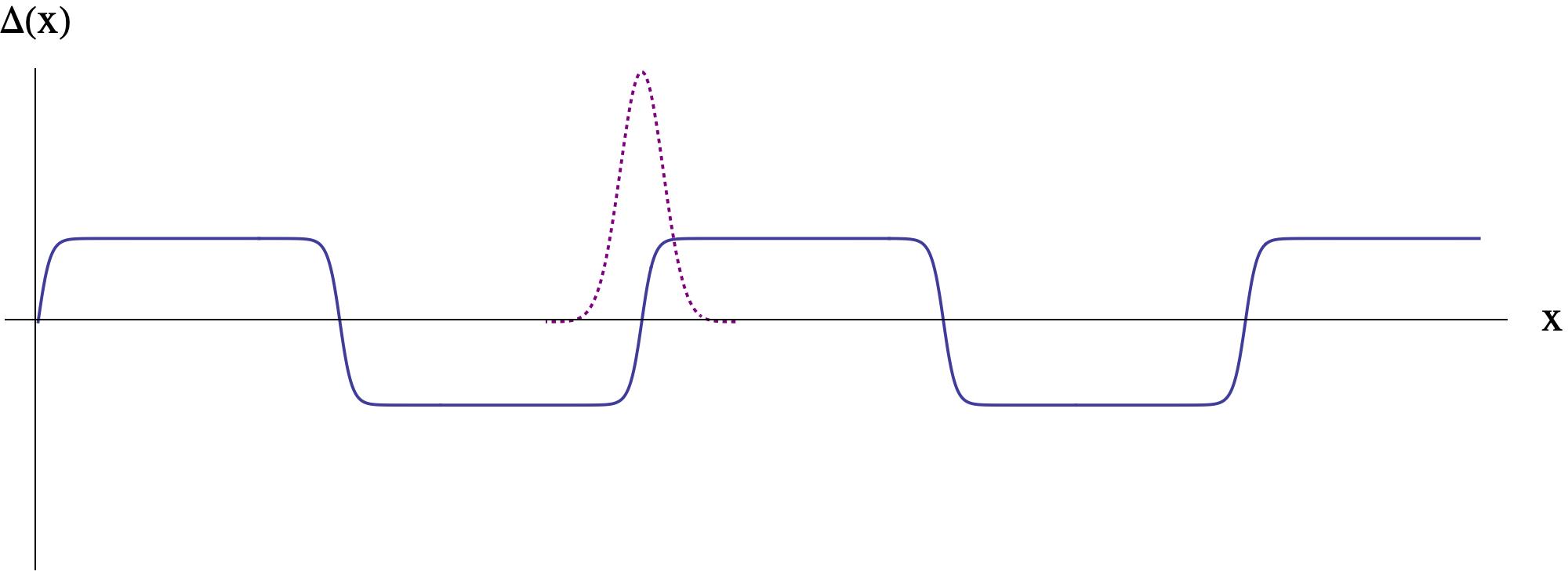} 
   \caption{A periodic snoidal wave. The dashed line represents a half-fermion zero mode localized on kinks.}
   \label{snoidal}
\end{figure}

In the limit of a large period, the snoidal wave \eq{7} degenerates to a  kink:
\begin{align}
  \Delta(x)\underset{k\to 1}{\to} \Delta_0\tanh \xi\,.
\label{vv}
\end{align}  
The formula 
\begin{align}
  {\rm sn}(\xi,k)=\frac{\pi}{2kK'}\sum_{n=-\infty}^\infty
(-1)^n \tanh \left(\frac{\pi}{2K'}(\xi+2nK)\right)
\label{13}
\end{align}
represents the traveling wave as a lattice of alternating bright-dark kinks, Fig.~\ref{snoidal}, where $K'=K(k')$ and $k'=\sqrt{1-k^2}$ is the complementary elliptic modulus. 
\subsection{The spectrum \la{1.2.2}}
We now consider a 2-gap solution of the Dirac equation \eq{6}. Among the extensive literature, we refer to papers on the commensurate Peierls phenomenon 
\cite{Bra_G_K1980,Bra_Dz_Kir_spin1981}. 

The states $\psi_E(x)$ in a periodic potential are characterized by the unitary Bloch factor $\Lambda(E)=e^{iPL}$, which defines the quasi-momentum as a multi-valued map $P(E)$. 
Symmetries of the Dirac operator imply that each value of $E$ corresponds to two momenta, $\pm P$, and each momentum $P$ corresponds to pairs $\pm E$ equal in number to the positive-energy bands. For a finite number of bands, the momentum is single-valued on a hyperelliptic Riemann surface. For two gaps, this surface is an elliptic curve determined by the spectrum endpoints $E_\pm$:
\begin{align}
 y^2 = (E_-^2 - E^2)(E_+^2 - E^2)\,.
 \la{21}
\end{align}
The spectrum consists of four bands, two of which merge at $E=0$ to form a central band:
$E:\ [-\infty,-E_+],\ [-E_-,E_-],\ [E_+,+\infty]$, with two gaps $[-E_+,-E_-],\ [E_-,E_+]$. The surface comprises two copies of the complex $E$-plane, glued along the bands. Fig.~\ref{spectrum} illustrates this spectrum.

The spectrum is fully characterized by the density of states:
\begin{align}
 \rho(E) =\frac{2\pi}{L}\left| \frac{dP}{dE}\right |\,,\la{}
\end{align} 
which is the central object of interest.

According to the general theory of finite-gap spectra \cite{AblowitzSegur}, the differential of the momentum is an Abelian differential on the complex curve \eq{21}:
\begin{align}
    dP= \frac{C-E^2 }{y(E)  } dE \,, \la{20}
\end{align}
where $(P,E)$ is considered as a complex point on the curve. The numerator in \eq{20} is a second-order polynomial, which in our case is even dues the reflection symmetry of the spectrum, with the leading coefficient fixed by the relativistic asymptote $|dP/dE|\to 1$ as $|E|\to \infty$.   
The constant $C$  determines   the density of states at $E=0$, where the kink's bands cross: $C={E_-E_+}\,|dP/dE|_{E=0}$. The physical spectrum corresponds to the real section of the spectral curve $P^2=F(E^2)$. 

The differential $dP$ has a second-order pole at $E=\infty$ with local parameter $E\sim 1/z$, making it an Abelian differential of the second kind.

It is subject to the normalization conditions:
\begin{align}
 &\int_{\rm bands} dP= (2\pi/L)\,
  N_s,\quad
   \int_{\rm gaps} dP=0\,.
\label{18}
\end{align}
(We recall that  $L/N_s$  is the period and $N_s$ is the number of bright (or dark) kinks, which also equals the number of fermions).  

The first integral is over the spectrum (bands) and the second over the gaps. In these terms, the conditions \eq{18} correspond to integrals over the $a$ and $b$ cycles of the elliptic curve \eq{21}: $\oint_b dP=(2\pi/L) \,N_s$, $\ \oint_a dP=0$. These normalization conditions express the spectrum endpoints $E_\pm$ and the constant $C$ in terms of $N_s/L$. Quoting the results \cite{Bra_G_K1980}:
\begin{align}
 &E_\pm = \frac{\Delta_0}{2} \left(k^{-1/2} \pm k^{1/2}\right),\quad
 C = E_+^2\frac{E(r ) }{K(r)},\quad r = 2(k^{1/2} + k^{-1/2})^{-1}
 \,.\la{2211}
\end{align}
Here the elliptic modulie $k$ is determined by the period of the wave by virtue \eq{131},
  \( r  \) is the Landen-transformed modulus, and \( E(r) \) is the complete elliptic integral of the second kind.
The algebro-geometric approach allows extension to multiple-gap solutions, albeit in a less explicit form  \cite{Bra_Dz_Kir_spin1981}.

In the paper, we obtain the density of states \eq{20} as the large-$N$ limit of the thermodynamic Bethe equations with $O(2N)$ symmetry relevant to the GN model.  
\\~\\
A short version of this work was published in \cite{MWZ}.  Other relevant references are:  on the semiclassical limit of thermodynamic Bethe equations and finite gap solutions; for KdV \cite{Smirnov:1998kv} and Landau-Lifshitz   \cite{Kazakov} equations;    \cite{Dunne} on relations between the GN-model and Peierls phenomenon.

\section{Finite-gap potentials and equilibrium quantum states \la{FG}}

Now we describe a direct correspondence between finite-gap solutions and a special class of eigenstates of the associated quantum field theory.

The spectrum of a relativistic integrable quantum field theory at zero chemical potential consists of a finite set of mass shells
\begin{align}
p^2=(p^0)^2-m_q^2,\quad q=1,\dots,N\,,\la{23}
\end{align}
each for a multiplet, where   $m_q$ is the particle mass spectrum. 
 In integrable systems, scattering of particles is factorized into a product of consecutive two-particle scattering events. We denote the scattering phase of a two-particle scattering by  $ \Theta_{qq'}(p,p')$.

Particles are in a semiclassical correspondence with solitons and their bound states in classical nonlinear equations.
By contrast, periodic solutions correspond to more intricate thermodynamic states. They involve a macroscopic number of particles in a given multiplet, ${N}_q\propto L$, scaled with the system size, whose momenta, denoted by \(P_q\), are piecewise uniformly distributed. Momentum is a multiply-valued function of energy \(E\). The differential of the spectrum map \(E\to P_q\), given by \(dP_q=(dP_q/dE)dE\), defines the density of states 
\be
\rho_q(E)dE:=\tfrac L{2\pi}|dP_q|,\quad \int\rho_q (E) dE=N_q\,.
\la{252}
\ee
The density of states is a defining, albeit local, characterization of the spectrum. 

At large energy, well exceeding the chemical potential, the spectrum tends toward the mass shells, hence  $|dP_q/dE|\to  1$.
This property motivates a parametrization of the spectrum by the mass shell as multiply-valued maps  $p\to P_q,\ p\to E_q$. 

A momentum $P_q$ differs from the momentum of the asymptotic state $p$ by scattering phases accumulated from interactions with other particles. The total scattering phase in a thermodynamic state is 
$\sum_{q'}\int   \Theta_{qq'}(p,p')\, \rho_{q'}(p') dp'$, where \(\rho_q(p)dp=\tfrac L{2\pi}|dP_q|\), but now $dP_q$ is the differential of the map $p\to P_q$. Hence, 
\begin{align}
P_q(p) = p + \frac 1{2\pi}\sum_{q'} \int   \Theta_{qq'}(p,p')\, |dP_{q'}(p')|\,. 
\label{251}
\end{align}
This is the thermodynamic Bethe equation, defining a differential of the map $p\to P_q$.

It is convenient to formulate the Bethe equations in terms of the \emph{rapidity} of asymptotic states, defined as the invariant relativistic measure
\begin{align}
d\theta = \left[\int_{-\infty}^\infty
\delta((p^0)^2-p^2-m_q^2)\, dp\right] dp^0 = \frac{dp_0}{\sqrt{(p^0)^2-m_q^2}}\,.\la{22}
\end{align}
Rapidity makes Lorentz boosts additive. Consequently, the two-particle scattering phase depends only on the difference of rapidities $\Theta_{qq'}(p,p')=\Theta_{qq'}(\theta-\theta')$. Viewed as a complex variable, rapidity uniformizes the mass shell:
\be
p^0 = m_q\cosh\theta,\quad p = m_q\sinh\theta\,. 
\ee
Differentiating the Bethe equation \eq{251}, we obtain an integral equation that determines the differential of the map \(\theta\to P_q\) 
\begin{align}
\tfrac{d}{d\theta} P_{q}(\theta)  = m_q\cosh\theta + \sum_{q'} \int   K_{qq'}(\theta-\theta')\, dP_{q'}(\theta')\,, 
\label{184}
\end{align}
where
\begin{align}
K_{qq'}(\theta) = \tfrac 1{2\pi}\tfrac{d}{d\theta}  \Theta_{qq'}(\theta)\,.
\end{align}

Similarly, the differential of energy $dE_q$ satisfies the relativistic counterpart of Eq.~\eqref{184}:
\begin{align}
\tfrac{d}{d\theta}E_{q}(\theta)  = m_q \sinh\theta+ \sum_{q'} \int   K_{qq'}(\theta-\theta')\, dE_{q'}(\theta')\,.
\label{205}
\end{align}
Once the scattering phases are known, the thermodynamic Bethe equations (\ref{184},\ref{205}) determine the differentials of the number of particles in each multiplet through the normalization of the momentum differential \eq{252}.

In Ref.~\cite{DNW} it was argued that solutions of these equations give a minimum to the total energy 
\begin{align}
\mathcal{E} =\sum_q \int E_q(p) \rho_q(p) dp=(L/2\pi)\sum_q\int E\, dP_q,
\end{align}
provided that the support of $\rho_q$ consists of a finite number of disjoint intervals.
These arguments prompt our {\it main proposition}:
\begin{mdframed}[linewidth=1pt]
The semiclassical limit of finite-interval thermodynamic Bethe states
yields the finite-gap spectrum of a linear operator (Lax operator) that appears as the
Hamiltonian in the adiabatic approximation of the underlying quantum model. If the system is relativistic, the Hamiltonian is the Dirac operator.
\end{mdframed} 
To support this statement we demonstrate that the differentials determined by the Bethe equations (\ref{184},\ref{205}) become the real section of an {\it Abelian differential of the second kind} single-valued on a Riemann surface determined by the endpoints of the spectrum. This is illustrated here on the example of the traveling wave.

The traveling wave we focus on corresponds to a {\it one-interva}l state filled by {\it equally distributed $O(2N)$ spinors of opposite chirality}. We denote the spectrum of the spinor state by $(P_s,\ E_s)$, and assume that spinor rapidities occupy an interval $|\theta|<B$, where $B$ is determined by the given number of spinors of the same chirality $N_s$ (in total $2N_s$ spinors) via condition \eq{252}. 
\VM{Then the support of the integrals in the Bethe equations is limited by the spinor mode
and set to be the interval $[-B, B]$. The Bethe equations are simplified further:}
\begin{align}
\frac{d}{d\theta}\begin{pmatrix}P_{s} \\
E_{s} \\
\end{pmatrix}  = m_{s}\begin{pmatrix}\cosh\theta \\
\sinh\theta\\
\end{pmatrix} +\, \int_{-B}^B K_{ss}(\theta-\theta')\,
d\begin{pmatrix}P_{s} \\
E_{s} \\
\end{pmatrix}, \label{31}
\end{align}
with
\begin{equation}
K_{ss}(\theta) = K_{\pm\pm}(\theta) + K_{\pm\mp}(\theta)\,.\la{33}
\end{equation}
Using the explicit form of the S-matrix yielding $K_{ss}$ (Sec.~\ref{SS},\ref{5.2}) and their large-rank limit (Sec.~\ref{6.1}), we recover the spectrum of the traveling wave (Sec.~\ref{7.3}) already described in Sec.~\ref{1.2.2}. We briefly comment on other states that yield the same semiclassical limit.

\section{Scattering by a kink: P\"oschl-Teller problem \la{PT}}
A building block of a periodic solution, and a benchmark of the underlying quantum theory, is the scattering phase of an elementary fermion by a kink, a potential given by \eq{vv}. It  is a special case of the classical  P\"oschl-Teller problem -- a Schr\"odinger operator \eq{1} with the  potential $u(x)=\Delta^2-\Delta_x=\Delta_0^2(1-2/\cosh^{2} \xi)$, \VM{where $\xi=\Delta_0x$}. 

In the Majorana basis, $\alpha=\sigma_2,\ \beta=\sigma_1$,  solution  of the Dirac equation \eq{6} with energy $p^0=\sqrt{\Delta_0^2+p^2}$ is  $\psi_p(x)=e^{ipx}\chi_p (x)$ with
\begin{align}
 \chi_p(x)
= \begin{pmatrix}{}
   \VM{({\Delta_0\tanh\xi-ip})/{p^0}}
   \\1
\end{pmatrix}\,.
\label{35}
\end{align}
At  $x\to\pm\infty$ the asymptotics of the wave function is given by the wave function of the Dirac operators with the uniform mass  $H_\pm^{(0)}= -i\alpha \partial_x \pm \beta \Delta_0$. They are connected by the chirality operator $\Gamma=\gamma^0\gamma^1=\sigma_2$ (a kink acts like chirality flip) $H_+^{(0)}=\Gamma H_-^{(0)}\Gamma$.  Hence,  
\begin{align}
 \Gamma \chi_p(+\infty)=A(p)\chi_p(-\infty)\,,\la{381}
\end{align}
where 
 \(  A(p)=\left(\frac{p+i\Delta_0}{p-i\Delta_0}\right)^{1/2}
\)
\VM{is a meromorphic function of rapidity equal to a pure phase at real $p=\Delta_0\sinh\theta$}. \VM{With $p=\Delta_0\sinh \theta$, $A$ is a meromorphic function} of rapidity
\begin{align}
A(\theta)=\tanh\left(\frac\theta 2+\frac{i\pi}4\right)\,,
\label{171}
\end{align}
having a physical pole $\theta=i\pi/2$. The pole  corresponds to a  Majorana zero-energy (counted from $\Delta_0$) bound state localized on the kink
\(
 \begin{pmatrix}
  1/{\cosh}\,  \xi\\
 0
 \end{pmatrix}
\). It is depicted in Fig.~\ref{snoidal}.
This is the Jackiw-Rebbi effect \cite{Jackiw} of emergence of a spinor state. As the number of kinks increases, they form a periodic structure, Fig.~\ref{snoidal}, and the zero-modes hybridize into the central (kink's) band, Fig.~\ref{spectrum}. 

We comment that the proper P\"oschl-Teller model describes only one chiral component of the Dirac multiplet, the transmission amplitude is a meromorphic function of the momentum $p$ equal to a square of  $A(p)$:\ 
\(A(p)/A(-p)=A^2(p)=\frac{p+i\Delta_0}{p-i\Delta_0}\).

In Sec.~\ref{5.2}, we show that $A(\theta)$ given by \eq{171} indeed arises as a semiclassical limit of the exact vector-spinor scattering amplitude in the GN model. 
Later in Sec.~\ref{5.4} we comment on how the complete quantum S-matrix has grown up from it.

\section{General properties of Lie group-invariant system} 

As we already stated, our system is an integrable system invariant under 
$O(2N)$ Lie group. The major property of Lie-group invariant is that the symmetry alone determines the major properties of the system.   We briefly review them following Ref. \cite{OW,ORW,RW}.  

 \subsection{Particle content \la{symmetry}}
The main properties of Lie group invariant systems are: 
\begin{itemize}
\item [-]  Particle content:    Particles are multiplets of isotopic subspaces of representations (generally reducible) which are in one-to-one correspondence with the fundamental representations. Hence, the number of distinct particles equals the rank of the Lie algebra. If the weight of a fundamental representation is minuscule \cite{FootnoteM},  the particle representation is irreducible.
\item [-]  The \(CT\) transformation: The charge conjugation and the time reversal that converts a particle to an antiparticle corresponds to a symmetry of the Dynkin diagram, that permutes the simple roots in a way that preserves the Cartan matrix and yields the outer automorphisms of the Lie algebra. 
\item[-]   Particles are bound states of "elementary" particles. They are a subset of minuscule representations. The Clebsch-Gordon decomposition of the Kronecker product of elementary representations contains other fundamental representations. The number of elementary particles is at most two in accordance with the degree of symmetry of the Dynkin diagram.

\end{itemize}
These properties have been established by direct Bethe-Ansatz solutions of $G$-integrable models. Here we accept them at the start. They and the factorized bootstrap fully determine the scattering S-matrix. 
\\~\\
In our case, the Dynkin scheme is \(D_N\). Representations of \(D_N\) are subject \({\rm mod}\ 4\)-periodicity.
This structure, however, does not show up in the semiclassical limit achieved at  \(N\to\infty\).  We, therefore, choose \(N/2\) to be an even integer when spinor representations are real-orthogonal. It helps to avoid unnecessary complications in formulas at finite $N$. 

The \(N\) fundamental representations of \(D_N\) are  \(N-2 \)  exterior powers \(\Lambda^r \vv,\ r=1,\dots, N-2\) of the defining representation \(\vv\) -- real \(2N\)-vectors,  and two  spinor representations \(s_\pm\)  with chirality $\pm$ of \({\rm Spin}\,(2N)\), each of dimension \(2^{N-1}\). 
 Among them, the vector and two spinors $v$ and $s_\pm$ are minuscule. The particle content of our  model, which we denote by \(V_q,\ q=1,\dots, N\), is isotopic subspaces of minuscule representations, and \(N-3\) reducible representations  generated by antisymmetric tensors \(\Lambda^r \vv\)
\be
V_q:\ \{V_r=\underset{p=0} {\overset {[r/2]}\oplus} \Lambda^{r-2p}\
 \vv, \ r=1,\dots,N-2,\quad  V_{N-1}=s_-,\quad  V_N=s_+\}\,.\la{181}
\ee

    The minuscule representations have a semiclassical meaning in the Peierls model  \eq{3}. Spinors correspond to the kinks \eq{vv},  and vectors are fermions in \eq{3}  \cite{Dashen}.   Other particles, the $N-3$ particles being bound states of vector particles, are not bound to their vector constituents. We, therefore,  focus on minuscule representations.
    
    The external automorphism,  the  {\it charge conjugation}, generated by the symmetry of the Dynkin diagram plays an important role. It is  
   a linear transformation defined by \(\mathbb{C}V_q=V_q^\ast\).   The conjugate representation, which we denote \(V_{\bar q}:=V_q^\ast\) is  a contravariant representation. It describes antiparticles.  $\mathbb{C}$ acts as an identity on representation other than spinors. At even \(N\), the charge conjugation keeps spinor chirality, so particles and antiparticle have the same chirality. If \(N\) is odd, the chirality of a spinor particle and its antiparticle are opposite. In symbols, if $\Gamma$ is  a chirality matrix ($\Gamma^2=1,\ \Gamma^{\rm T}=\Gamma$), the $\Gamma\mathbb{C}=(-1)^N\mathbb{C}\Gamma$.  Furthermore, if $N/2$ is an even integer, the case we choose, the charge conjugation matrix $\mathbb{C}$ is symmetric (in general $\mathbb{C}^{\rm T}=(-1)^{[N/2]}\mathbb{C}$. Finaly, the many-body states we consider consist of an even number of spinor particles (avoiding another source of alternating signs, this time in eigenvalues of the monodromy matrix).

 \subsection{Factorized theory of scattering\la{B}}
   \subsubsection{Factorized S-matrix}
   In integrable models, the scattering is factorized into a consecutive associative product of two-particle scattering.   The scattering matrix \(S_{qq'  }(p,p')\), the building blocks of the theory, 
  is a function of momenta and representations of scattered particles. The Lorentz invariance yields that the S-matrix depends on the difference of rapidities \(S_ {qq'  }(\theta-\theta')\). It acts in the isotopic subspaces of the product of representations of each particle \(V_q\otimes V_{q'}\). 
     
     The  scattering is factorized if the associativity condition for the scattering of three particles, referred to as  the Yang--Baxter equation, holds:
 \begin{align}
 S_  {qq'  }(\theta)S_  {qq''  }(\theta+\theta')S_  {q'q''  }(\theta')=\VM{S_  {q'q''  }(\theta')}S_  {qq''  }(\theta+\theta')S_  {qq'  }(\theta)\,.\la{YB}
\end{align}

\subsubsection{Cross-unitarity}
The  particle-antiparticle scattering matrix acts in \(V_{\bar q}\otimes V_{q'}\). It is obtained by applying charge conjugation together with the sign reversal of energy:
         \begin{align}S_  {\bar qq'  }(\theta)=\mathbb{C}\, S_  {qq'  }(\theta^C)\, \mathbb{C}\,,\la{crossing}\end{align}
  where \(\mathbb{C}\) acts in the space of the first particle $V_q$  and 
  \begin{align}
  \theta^C=i\pi-\theta\,.
\end{align}
  The particle--particle and particle--antiparticle S-matrices are both unitary:
  \begin{align} 
  S_  {qq'  }(\theta)S_  {qq'  }(-\theta)=1,
 \quad S_  {\bar qq'  }(\theta)S_  {\bar qq'  }(-\theta)=1\,.\la{U}
\end{align} 
 \subsubsection{Spectral decomposition}
 The eigenvalues of the S-matrix correspond to the Clebsch-Gordan decomposition of the tensor product of the fundamental representations into a sum of irreducible components:
\begin{equation}
  \vv_q\otimes \vv_{q'}=\underset{\omega_{qq'}} \oplus\  \vv_{\omega_{qq'}}\,,\la{254}
\end{equation}
where $\omega_{qq'}$ enumerates the irreducible representations in the decomposition. We use $\omega$ to denote the highest weight of the representation.
\VM{Then it follows from Schur's lemma that}
\begin{equation}
  S_  {qq'  }(\theta)=\sum_{\omega_{qq'}} s_{\omega_{qq'}}(\theta)\, \mathscr{P}_{\omega_{qq'}}\,,\la{26}
\end{equation}
where $\mathscr{P}_\omega$ is the projector onto the space of weight $\omega$. The amplitudes $s_{\omega}(\theta)$ are unitary 
\begin{equation}
  s_\omega(\theta)\,s_\omega(-\theta)=1
\end{equation}
  meromorphic functions with at most simple poles. They are the eigenvalues of the S-matrix.

    \subsubsection{Bound states and  bootstrap}  
 Eqs. \eq{YB} and \eq{U}   determine the S-matrix up to a real scalar factor obeying the cross-unitarity condition
 \begin{align}
 \mathfrak{X}_  {qq'  }(\theta)\,\mathfrak{X}_  {qq'  }(-\theta)=1,\quad \mathfrak{X}_  {\bar qq'  }(\theta)\,\mathfrak{X}_  {\bar qq'  }(-\theta)=1\,,\la{241}
\end{align}
where \(\mathfrak{X}_  {\bar qq'  }(\theta)=\mathfrak{X}_  {qq'  }(i\pi-\theta)\).
 Solutions  of \eq{241} are characterized by a finite set of parameters \(\theta_  {qq',j}\) representing poles on the physical sheet:
 \begin{align}
 \mathfrak{X}_  {qq'  }(\theta)=\VM{\prod_j\frac{\sinh\left( \tfrac 12(\theta+\theta_{qq',j})\right) }{\sinh\left(\tfrac 12(\theta-\theta_ {qq',j  })\right) }\,.}\la{CDD}
\end{align}
 Some of the poles,  usually purely imaginary,  correspond to bound states, i.e., additional particles. The positions of the physical poles determine the  
  mass of the bound state \footnote{See \cite{ORW} for a sufficient condition for a pole to represent a bound state}. We therefore introduce a minimal solution of (\ref{YB},\ref{U}) that is analytic on the physical sheet, denoted \(\mathfrak{S}(\theta)\). Then the S-matrix can be written as the product of this matrix and the scalar factor:
  \begin{align}S_  {qq'  }(\theta)=\mathfrak{X}_  {qq'  }(\theta)\ \mathfrak{S}_{qq'  }(\theta)\,.\la{242}\end{align}
The matrix factor represents kinematic and symmetry properties and may be common to a class of models, whereas the scalar factor encodes model-specific dynamics.

These are the elements of the bootstrap strategy, summarized as follows.   We seek only the S-matrix of the elementary particles, which in our case are spinors
 all other particles appear as bound states.
 The S-matrix of the elementary particles is obtained in two steps. The first step is to find a {\it minimal} solution  \(\mathfrak{S}(\theta)\) of the Yang--Baxter equation and cross-unitarity. 
It satisfies all kinematic constraints. The model-specific dynamics are encoded in the positions of the poles of the factor \(\mathfrak{X}\) in \eq{CDD}. In $G$-invariant systems, the poles are introduced so that the spectrum of bound states is minimal: bound states of bound states must not extend the spectrum beyond the minimal particle set. In $G$-invariant systems, this minimal content consists of the fundamental representations, as discussed in Sec.~\ref{symmetry}.

    \subsubsection{Fusion formula}  

If a physical pole $ \vartheta_b$ appears in the amplitude $s_{q_b}$ of the spectral decomposition of $S_{qq'}$, a bound state is the multiplet of the representation $V_{q_b}$. The {\it fusion formula} \cite{K0,ORW} gives the scattering matrix of the $V_{q_b}$-particle. If $q$ is a minuscule representation, the pole appears in a single channel. In this case, the fusion formula simplifies:
\begin{align}
  S_{q_b,q''}(\theta)=\mathscr{P}_{q_b}\left[S_  {qq''  }(\theta+  \tfrac 12\vartheta_b)S_  {q', q''  }(\theta- \tfrac 12\vartheta_b)\right]\mathscr{P}_{q_b}\,.\la{F}
\end{align}

A pole $\vartheta_b$ that gives rise to a bound state must lie on the physical sheet, and its residue has to satisfy a specific condition \cite{K0}. For minuscule representations, this reduces to the sign of the residue of the factor \(X\):  \({\rm res}\, X_{qq'}<0\). Then the mass of the bound state is expressed in terms of the masses of the scattered particles and the position of the pole. \VM{ If the parent particles have the  same mass $m$, the mass of the bound state is}
  \begin{align}
m_b=2m\cosh \left( \tfrac 12 {\vartheta_b}\right)\,.\end{align}

\subsubsection{Cartan component}
 A general result of representation theory states that among all highest weights  appearing in the Clebsch--Gordan decomposition of representations with highest weights \(\omega_q\) and \(\omega_{q'}\), the maximal dominant weight is the sum: 
 \begin{align}\Omega_{qq'}=\omega_q+\omega_{q'}\,.\end{align}
 This component is referred to as the Cartan component of the tensor product, or the top Clebsch--Gordan component. \VM{We denote it by $\Omega$}. It appears with multiplicity one. The highest weights of the remaining representations in the decomposition \eq{254} are obtained from the top component by subtracting multiples of simple roots: 
 \( \omega=\Omega-\sum_i m_i\alpha_i\).  
 The amplitudes of the Cartan components serve as a kind of "root" for (nested) Bethe Ansatz equations.   The scattering phases in Eq.~(\ref{31}) of Sec.~\ref{FG} correspond to the Cartan components
 thus, they are the only components we need.
 Furthermore, for Cartan components, the fusion formula \eq{F} becomes multiplicative:
 \begin{align}
  s_{\Omega_{b}}(\theta)=s_{\Omega_{q}}(\theta+\tfrac 12\vartheta_b)\, s_{\Omega_{q'}}(\theta-\tfrac 12\vartheta_b)\,.
\label{49}
\end{align}


The Bethe equations (\ref{251}) equate the Bloch factors \(e^{iP_qL}\) to the eigenvalues of the monodromy matrix (also called the transfer matrix), given  by the product of scattering matrices of a particle \(q\) with all other particles labeled by \(q_i\):
\begin{align}
 T_q(\theta,
\{\theta_i\})=e^{ip L}\prod_{i} S_  {qq_i  }(\theta-  \theta_i),\quad p=m_{q}\VM{\sinh}\theta \,.
\end{align}
The monodromy matrix acts in the space
\begin{align}V_{q}\otimes (\underset{i}\otimes V_{q_i})=\underset{\omega}\oplus V_{\omega}\,,\la{291}\end{align}
where $\omega$ labels the highest weights of the irreducible representations appearing in the decomposition.
 Accordingly, we decompose the monodromy matrix as 
\begin{align}
T_q(\theta,
\{\theta_i\})=e^{ipL}\sum_\omega t_\omega(\theta,
\{\theta_i\})\mathscr{P}_ {\omega  }\,.
\end{align}

Among the states in \eq{291}, we are interested in the Cartan component, the state with the maximal highest weight: 
\begin{equation}
  \Omega=\VM{\omega_q + \sum_i \omega_{q_i}\,}.
\end{equation}
For this component, the eigenvalues of the monodromy matrix reduce to the product of the Cartan amplitudes of the two-particle S-matrices in \eq{26}:
\begin{equation}
  t_{\Omega}(\theta,\{\theta_i\})=\prod_{i} 
  s_{\Omega_q+\Omega_{q_i}}(\theta-  \theta_i)\,.
\end{equation}
Hence the kernels in Eqs.~(\ref{31}) are 
\begin{align}K_{qq_i}(\theta)=\frac 1{2\pi i}\frac d{d\theta}\log s_{\Omega_q+\Omega_{q_i}}(\theta)\,,
\end{align}
where $q$ is a spinor or vector representation and $q_i$ are spinor representations.
In the next section, we review the explicit forms of $K_{qq_i}$.   

  \section{O(2N)-scattering matrix \la{S}}
Here we review the $O(2N)$ scattering matrix. The vector-vector and vector-tensor components of the S-matrices were computed in Ref.~\cite{ZZ}, and the complete S-matrix, including vector-spinor and spinor-spinor components, was given in Ref.~\cite{Karowski}. We follow Refs.~\cite{OW,ORW, RW,R}, some formulas in this section are new.

 \subsection{Spinor-spinor scattering \la{SS}}
  The spinor S-matrix acts in the representation spaces  \(s_\pm\otimes s_\pm,\ s_\pm\otimes s_\mp\). Their  Clebsch-Gordon decompositions  (\(N={\rm even}\)) are
 \begin{align}
 s_\pm\otimes s_\pm=\Lambda^{0}\, \vv\oplus \Lambda^{2}\, \vv\oplus\dots\oplus \Lambda_\pm ^N \vv,\quad\quad s_+\otimes s_-=\Lambda^{1}\, \vv\oplus \Lambda^{3}\, \vv\oplus\dots\oplus \Lambda^{N-1}\, \vv.\,\la{30}
\end{align}
 In terms of highest weights (expressed in the usual orthonormal basis $e_1,\dots, e_N$) 
        $\mathrm{hw}\big(\Lambda^{0}\vv\big)=0$ and $ \mathrm{hw}
        \big(\Lambda^{r}\vv\big)=e_1+e_2+\cdots+e_r$ if $r<N$. 
        For even \(N\), the case we focus on, \(\Lambda^N \vv\) is a reducible representation which splits in two parts \(\Lambda^N_\pm \vv\), each belongs to \(s_\pm\otimes s_\pm \). Their  $\mathrm{hw}\big(\Lambda^{N}_{\pm}\vv\big)= e_1+e_2+\cdots+e_{N-1}\pm e_N$. The spinors $\mathrm{hw}\big(s_\pm\big)=\tfrac 12( e_1+e_2+\cdots+e_{\,N-1}\pm e_N)$.
  The Cartan components of  $s_+\otimes s_-$ and $s_\pm\otimes s_\pm$  are the largest even/odd-rank antisymmetric tensors  \(\Lambda^{N-1} \vv\) and \(\Lambda_\pm ^N \vv\).

According to the decomposition \eq{30} and \eq{26} the  S-matrices of spinors with like/opposite chirality \(S_  {\pm \pm  }\)  and \(S_  {\pm \mp  }\) are  given  by the formula
 \begin{align}
S_{\pm \pm }(\theta)\pm S_{\pm \mp }(\theta)= \sum_{q=0}^{N}(\pm 1)^q s_{q}(\theta) \mathscr{P}_  q   \la{25}
\end{align}
with unitary  amplitudes 
\begin{align}
 s_q(\theta)s_q(-\theta)=1\,.
\end{align}
The Yang-Baxter equation \eq{YB} establishes the relation between the amplitudes
\begin{align}
 s_{N-q-1}(\theta)=(-1)^q\frac{\theta+\frac{2\pi i}{ h^\vee} q}{\theta-\frac{2\pi i}{ h^\vee}q}\ s_{N-q+1}(\theta)\,,\la{261}
\end{align}
where \begin{align}\text{dual Coxeter number}:\qquad h^\vee=2N-2\end{align}is the $D_N$ dual Coxeter number (the dual Coxeter number and the Coxeter number are equal for the simply-laced root systems).
This leaves two amplitudes undetermined. We choose them to be  Cartan components  denoted  by  $s_{N-1}:=s_{+-}=s_{-+}$ and $s_N:=s_{++}=s_{--}$.

The amplitudes are further restricted by the cross-unitarity conditions. They are based on the action of the charge conjugation $\mathbb{C}$  in \eq{crossing}. For $N/2$ being an even integer 
\begin{align}
&s_{\pm\pm}(\theta)=-
\prod_{0\leq q={\rm even}}^{N}\frac{\theta^C-\frac{2\pi i }{ h^\vee} q}{\ \  \theta-\frac{2\pi i}{ h^\vee}q}\ {s}^C_{\pm\pm}(\theta),\ 
&s_{\pm\mp}(\theta)=
\prod_{1\leq q={\rm odd}}^{N-1}\frac{\theta^C-\frac{2\pi i}{ h^\vee}q}{ \ \ \, \theta-\frac{2\pi i}{ h^\vee}q}\ s^C_{\pm\mp}(\theta)\,,\la{C}
 \end{align}
 where we denoted $s^C(\theta)=s(\theta^C)$.
 
These data and the minimal pole assumption are sufficient to find the amplitudes   \(s_q(\theta)\). In particular, the Cartan components follow solely from cross-unitarity and the pole assumption. They are
\begin{align}
 \log s_{++}(\theta)\pm \log s_{+-}(\theta) =
\sum_{r=0}^{N-2}(\pm 1)^r\ 
\log \frac{\Gamma\left(\tfrac 12- \frac{\vartheta_r+\theta}{2\pi i} \right)\Gamma\left(\tfrac 12+ \frac{\vartheta_{r+1}-\theta}{2\pi i}  \right) }{\Gamma\left( \tfrac 12- \frac{\vartheta_r-\theta}{2\pi i}\right)\Gamma\left(\tfrac 12+ \frac{\vartheta_{r+1}+\theta}{2\pi i}  \right) }\,,\la{68}
\end{align}
where $\vartheta_r$ is given by \eq{281} below.

\VM{For references we give here the factor  \(\mathfrak{X}_{ss}\), which introduces the physical poles,   and the factors \(\mathfrak{S}_N:=\mathfrak{S}_{++},\ \mathfrak{S}_{N-1}:=\mathfrak{S}_{+-}\) - the minimal solutions of the Yang-Baxter equations that do not contain physical poles. We recall their definitions   \(s_q(\theta)=(-1)^q\mathfrak{X}_{ss}(\theta)\mathfrak{S}_q(\theta)\)  }
\begin{align}
\mathfrak{X}_{ss}(\theta)=\prod_{r=0}^{N-2}\frac{\sinh\tfrac 12(\theta+\vartheta_r)
}
{
\sinh \tfrac 12(\theta-\vartheta_r)
}\,,\la{38}
\end{align}
\begin{align}
\log \mathfrak{S}_{++}(\theta)\pm \log \mathfrak{S}_{+-}(\theta)=\sum_{r=0}^{N-2}(\pm  1)^r \
 \log\frac{\Gamma\left(\tfrac 12+ \frac{\vartheta_r-\theta}{2\pi i}  \right)
\Gamma\left(\frac 12-\frac{\vartheta_{r+1}-\theta}{2\pi i}\right)
}
{\Gamma\left(\tfrac 12+ \frac{\vartheta_r+\theta}{2\pi i}   \right)
\Gamma\left( \frac 12-\frac{\vartheta_{r+1}+\theta}{2\pi i} \right)
}\,.
 \end{align}
\VM{The  physical poles }
\begin{align}  \vartheta_r={i\pi}\left(1-{2r}/{h^\vee}\right),\quad \vartheta^C_r=\vartheta_{N-r-1},\quad  r=\VM{1,\ldots,  N-2\,\la{281}}
\end{align}
appear in the channels of representations entering  $V_r$, see  \eq{181}.
They correspond to  bound states with the mass given by
   \begin{equation}
 m_r=
 2m_s\sin(\pi r/h^\vee),\quad  r=1,\ldots,  N-2\,,\la{191}
 \end{equation}
 where $m_s$ is the mass of the spinor.
There is no bound state in the scalar channel and in the Cartan components. This is the effect of the bootstrap: bound states do not proliferate beyond the Dynkin diagram.

%

Summing up,  the scattering phases of Cartan components $\Theta_{\pm\pm}=\frac 1{ i }  \log s_N,\ \Theta_{\pm\mp}=\frac 1{ i }  \log s_{N-1}$, and their derivatives are now given
\begin{align}
 K_{\pm {\pm}}=\frac 1{2\pi }\frac d{d\theta}  \Theta_{\pm\pm},\quad K_{\pm {\mp}}=\frac 1{2\pi }\frac d{d\theta}  \Theta_{\pm\mp}\,.\la{48}
\end{align}
They are  compactly written in terms of the Fourier integral that is also valid for $N$ odd or even, alike
\be
\begin{aligned}
 K_{\pm\pm}(\theta)=\delta(\theta)+&\tfrac 1{4\pi}\int_{0}^\infty  
\left(
\frac{\tanh \frac{\pi x}{2}}{e^{-\frac{2\pi x}{ h^\vee}}-1}-
\frac{1}{e^{-\frac{2\pi x}{ h^\vee}}+1}\right)\cos( x\theta)d x\,,\la{46}\\
K_{\pm\mp}(\theta)=&\tfrac 1{4\pi}\int_{0}^\infty  
\left(
\frac{\tanh \frac{\pi x}{2}}{e^{-\frac{2\pi x}{ h^\vee}}-1}+
\frac{1 }{e^{-\frac{2\pi x}{ h^\vee}}+1}\right)\cos( x\theta)d x\,.
\end{aligned}
\ee
For references, we mention the Fourier transform of the factor \eq{38} \(\mathfrak{X}_{ss}=-\tfrac 1{8\pi}\int_{0}^\infty   
\frac {\tanh \left(\frac{\pi x}{2}\right)}{\sinh\left(\frac{\pi x}{ \ h^\vee}\right)}{\,\cos (x\theta)
d x}\). 
Then $K_{ss}(\theta)$, defined by \eq{33} \begin{align}
K_{ss}(\theta):=K_{\pm \pm}(\theta)+K_{\pm \mp}(\theta)=\delta(\theta)+\tilde K_{ss}(\theta)\,,\la{K}
\end{align}
with \begin{align}
\tilde K_{ss}(\theta)=\tfrac 1{2\pi}\int_{0}^\infty  
\frac {\tanh \frac{\pi x}{2}}{e^{-\frac{2\pi x}{ h^\vee}}-1}{\,\cos (x\theta)
d x}\,.\la{73}
\end{align}
\subsection{Fermion-spinor scattering \la{5.2}}
Elementary fermions form the vector multiplet \(v\).  For \(N\) even, they can be viewed as bound states of spinors with opposite chirality. We focus on their scattering with a spinor, the quantum analog of the P\"oschl
-Teller problem (Sec.~\ref{PT}).  There are only two channels:
\begin{equation}
\vv \otimes s_{\pm} = s_{\mp} \oplus (3/2)_{\pm}\,,
\end{equation}
 corresponding to a spinor with opposite chirality and the Cartan component: a spin-\(3/2\) Rarita--Schwinger field with
\(\mathrm{hw}(3/2_\pm) = \tfrac{3}{2} e_1 + \tfrac{1}{2} (e_2 + \cdots \pm e_N)\).

The Yang-Baxter equation relates the two  amplitudes, giving the vector-vector S-matrix
\begin{equation}
S_{\vv\pm}(\theta) = s_{\vv\pm}(\theta)\left(
\frac{\theta + \tfrac{1}{2} \vartheta_1}{\theta - \tfrac{1}{2} \vartheta_1} \mathscr{P}_{\mp} + \mathscr{P}_{3/2_\pm}
\right)\,.\la{701}
\end{equation}
Here \(\vartheta_1\) is given in \eq{281}, and \(\mathscr{P}_{\pm}, \mathscr{P}_{3/2_\pm}\) are projectors onto spinor and spin-3/2 representations with chirality \(\pm\).

The cross-unitarity 
\begin{align}
s_{\vv \pm}^C(\theta) = s_{\vv \pm}(\theta)\, \frac{\theta - \tfrac{1}{2} \vartheta_1}{\theta^C - \tfrac{1}{2} \vartheta_1}\,,
\end{align}
 which yields the minimal solution
\begin{align}
\mathfrak{S}_{\vv s}(\theta) =
\frac{
\Gamma\left(\frac{1}{2} + \frac{\theta - \tfrac{1}{2} \vartheta_1}{2\pi i}\right)
\Gamma\left(1 - \frac{\theta + \tfrac{1}{2} \vartheta_1}{2\pi i}\right)
}{
\Gamma\left(\frac{1}{2} - \frac{\theta + \tfrac{1}{2} \vartheta_1}{2\pi i}\right)
\Gamma\left(1 + \frac{\theta - \tfrac{1}{2} \vartheta_1}{2\pi i}\right)
}\,.\la{80}
\end{align}

The pole assignment ensures that the bound state of a vector and a spinor is a spinor (as indicated by the 
P\"oschl
-Teller problem):
\begin{align}
\mathfrak{X}_{\vv \pm}(\theta) = \pm \frac{\tanh \frac{1}{2} (\theta + \tfrac{1}{2} \vartheta_1)}{\tanh \frac{1}{2} (\theta - \tfrac{1}{2} \vartheta_1)}\,.\la{77}
\end{align}
\footnote{Sign factors of the amplitudes follow from the fusion formula \eq{F}.}

Collecting terms:
\begin{align}
s_{\vv\pm}(\theta) \equiv \mathfrak{X}_{\vv\pm}(\theta) \mathfrak{S}_{\vv s}(\theta)
= \pm \frac{
\Gamma\left(\frac{-\theta + \tfrac{1}{2} \vartheta_1}{2\pi i}\right)
\Gamma\left(\frac{1}{2} + \frac{\theta + \tfrac{1}{2} \vartheta_1}{2\pi i}\right)
}{
\Gamma\left(\frac{\theta - \tfrac{1}{2} \vartheta_1}{2\pi i}\right)
\Gamma\left(\frac{1}{2} - \frac{\theta + \tfrac{1}{2} \vartheta_1}{2\pi i}\right)
}\,.\la{79}
\end{align}

The Fourier representations of the vector-spinor  and  rank--\(r\) antisymmetric tensor- spinor are
\begin{align}
K_{\vv s}(\theta) := \frac{1}{2\pi i} \frac{d}{d\theta} \log s_{\vv\pm}(\theta)
= -\frac{1}{\pi} \int_0^\infty
\frac{ e^{\frac{\pi x}{\ h^\vee}}}{\cosh \frac{\pi x}{2}} \cos(x\theta) \, dx\,.\la{771}
\end{align}
\begin{align}
K_{rs}(\theta) = \sum_{n=-(r-1)/2}^{(r-1)/2} K_{\vv s}\left(\theta + i \frac{2 n \pi}{\ h^\vee}\right)
= -\frac{1}{\pi} \int_0^\infty
\frac{ e^{\frac{\pi x}{\ h^\vee}} }{\cosh \frac{\pi x}{2}}
\frac{\sinh \frac{\pi x}{\ h^\vee} r}{\sinh \frac{\pi x}{\ h^\vee}} \cos(x\theta) \, dx\,.\la{84}
\end{align}

\subsection{Fermion-fermion scattering \la{FF}}
Elementary fermions form vector multiplets of \(O(2N)\). Their scattering possesses  three channels: singlet, antisymmetric rank-2 tensor, and traceless symmetric rank-2 tensor:
\begin{equation}
\vv \otimes \vv = S^2 \vv \oplus \Lambda^2 \vv \oplus \Lambda^0 \vv\,.
\end{equation}
Accordingly, the S-matrix reads
\begin{align}
S_{\vv\vv}(\theta) = s_{\vv\vv}(\theta) \Bigg(
\mathscr{P}_S
+ \frac{ \theta  + \frac{2\pi i}{\ h^\vee}}{ \theta   - \frac{2\pi i}{\ h^\vee}} \mathscr{P}_A
+ \frac{ \theta    + \frac{2\pi i}{\ h^\vee}}{ \theta- \frac{2\pi i}{\ h^\vee} }\ \frac{\theta + i \pi}{\theta - i \pi} \mathscr{P}_0
\Bigg)\,,\la{76}
\end{align}
where   \(\mathscr{P}_S, \mathscr{P}_A, \mathscr{P}_0\) are the corresponding projectors and the Cartan component is
\begin{align}
s_{\vv\vv}(\theta) = \mathfrak{X}_{\vv\vv}(\theta) \mathfrak{S}_{\vv\vv}(\theta)
\end{align}
with
\begin{align}
\mathfrak{S}_{\vv\vv}(\theta) =
\frac{
\Gamma\!\left( \frac{-\theta}{2\pi i}+1 \right)
\Gamma\!\left( \frac{\theta}{2\pi i} + \frac{1}{2} \right)
\Gamma\!\left( \frac{\vartheta_1^C+\theta}{2\pi i} \right)
\Gamma\!\left( \frac{\vartheta_1^C-\theta}{2\pi i} + \frac{1}{2} \right)
}{
\Gamma\!\left( \frac{\theta}{2\pi i} +1\right)
\Gamma\!\left( \frac{-\theta}{2\pi i} + \frac{1}{2} \right)
\Gamma\!\left( \frac{\vartheta_1^C-\theta}{2\pi i} \right)
\Gamma\!\left( \frac{\vartheta_1^C+\theta}{2\pi i} + \frac{1}{2} \right)
},\quad
\mathfrak{X}_{\vv\vv}(\theta)= -\frac{\tanh \frac{1}{2} (\theta + \vartheta_1)}{\tanh \frac{1}{2} (\theta - \vartheta_1)}\,.\la{83}
\end{align}

The Fourier transform of the vector-vector scattering is
\begin{align}
K_{\vv\vv} (\theta)= \frac{1}{2\pi i} \frac{d}{d\theta} \log s_{\vv\vv}(\theta)
= \delta(\theta) + \tilde K_{\vv\vv} (\theta),\quad  \tilde K_{\vv\vv}(\theta)=\frac{1}{\pi} \int_0^\infty \frac{\cos \frac{x \vartheta_1}{2}}{\cosh \frac{\pi x}{2}} e^{\frac{\pi x}{\ h^\vee}} \cos(x\theta) dx\,.\la{82}
\end{align}
Similarly Fourier transform of the tensor-tensor scattering are  $K_{rr'}=K_{r'r}= \delta_{rr'} \delta(\theta)+\tilde K_{rr'} (\theta)$ with
\begin{align}
\tilde K_{rr'} (\theta)=- \frac{1}{\pi} \int_0^\infty  \frac{\cos \frac{x}{2} \vartheta_r}{\cosh \frac{\pi x}{2}} \frac{\sin \frac{x}{2} \vartheta^C_{r'}}{\sin \frac{x}{2} \vartheta^C_1} e^{\frac{\pi x}{\ h^\vee}} \cos(x\theta) dx,\quad r' \ge r\,.\la{88}
\end{align}

\subsection{Fusion \la{5.4}}
Scattering amplitudes satisfy fusion relations \eq{F}. They, together with a minimal analytic assignments, allow all amplitudes to be determined from a single known amplitude, e.g., \(s_{\vv\vv}(\theta)\), and for that reason could be used as defining conditions for the bootstrap approach.

A pole \(\vartheta_1\) in spinor-spinor amplitude indicates that for even \(N\) an elementary fermion can be viewed as a bound state of a spinor and its antiparticle of opposite chirality. 

The fusion formula \eq{49} for Cartan components  then yields:
\begin{align}
s_{++}\left(\theta - \tfrac{1}{2} \vartheta_1\right) s^C_{+-}\left(\theta + \tfrac{1}{2} \vartheta_1\right)
= \pm s_{\vv\pm}(\theta),\quad
\text{or}\quad
\frac{s_{++}\left(\theta + \frac{i \pi}{\ h^\vee}\right)}{s_{+-}\left(\theta - \frac{i \pi}{\ h^\vee}\right)} = \pm s_{\vv\pm}\left(\theta +\frac{i \pi}{2}\right)\,.\la{90}
\end{align}

Similarly, vector-vector fusion gives
\begin{align}
s_{\vv\pm}\left(\theta - \tfrac{1}{2} \vartheta_1\right) s_{\vv\mp}\left(\theta + \tfrac{1}{2} \vartheta_1\right)
= s_{\vv\vv}(\theta)\,.\la{91}
\end{align}

It is important to notice  that  fusion relation holds for $\mathfrak{X}$-factors and  the minimal solution $\mathfrak{S}$ separately:
\begin{align}
-\frac{\mathfrak{X}_{ss}\left(\theta + \frac{i \pi}{\ h^\vee}\right)}{\mathfrak{X}_{ss}\left(\theta - \frac{i \pi}{\ h^\vee}\right)} = \pm \mathfrak{X}_{\vv\pm}\left(\theta +\frac{i \pi}{2}\right),\qquad 
\frac{\mathfrak{S}_{++}\left(\theta + \frac{i \pi}{\ h^\vee}\right)}{\mathfrak{S}_{+-}\left(\theta - \frac{i \pi}{\ h^\vee}\right)} =  \mathfrak{S}_{\vv s}\left(\theta +\frac{i \pi}{2}\right)\,.\la{903}
\end{align}
\begin{align}
\mathfrak{X}_{\vv\pm}\left(\theta - \tfrac{1}{2} \vartheta_1\right) \mathfrak{X}_{\vv\mp}\left(\theta + \tfrac{1}{2} \vartheta_1\right)
= \mathfrak{X}_{\vv\vv}(\theta),\quad \mathfrak{S}_{\vv\pm}\left(\theta - \tfrac{1}{2} \vartheta_1\right) \mathfrak{S}_{\vv\mp}\left(\theta + \tfrac{1}{2} \vartheta_1\right)
= \mathfrak{S}_{\vv\vv}(\theta)\,.\la{913}
\end{align}

The  fusion relations imply relations for the differentials of the scattering phase
\begin{align}
 \tilde K_{\pm\pm}\left(\theta + \frac{i \pi}{\ h^\vee}\right) - K_{\pm\mp}\left(\theta - \frac{i \pi}{\ h^\vee}\right) &= K_{\vv s}\left(\theta + \frac{i \pi}{2}\right)\,,\la{93}\\
K_{\vv s}\left(\theta + \frac 12\vartheta_1\right) + K_{\vv s}\left(\theta - \frac 12\vartheta_1\right) &= \tilde K_{\vv\vv}\left(\theta\right)\,.\la{891}
\end{align}
and the relation
\begin{align}
K_{ss}\left(\theta - \frac{2i \pi}{\ h^\vee}\right) - K_{ss}\left(\theta + \frac{2i \pi}{\ h^\vee}\right) =
\sum_{\varepsilon=\pm}\left[K_{\vv s}\left(\theta -\frac{i\pi}2+\varepsilon \frac{i \pi}{\ h^\vee}\right)
-K_{\vv s}\left(\theta +\frac{i\pi}2+\varepsilon \frac{i \pi}{\ h^\vee}\right)\right]\,.\la{901} 
\end{align}
followed from(\ref{93}) . 
Integral representations (\ref{48},\ref{K},\ref{771},\ref{82}) provide an immediate check. 
\\~\\
The fusion rules are powerful relations reflecting no more that  the structure of the scheme \(D_N\). They could be seen as defining conditions for the bootstrap: knowing \(s_{v\pm}(\theta)\), all other amplitudes are determined by the fusion.

\section{Semiclassical limit of scattering \la{6.1}}
Now we are armed to study the limit $h^\vee\to\infty$. 

We recall that we focus on a thermodynamic state that consists of a large number $N_s$ of spinors with $+$ chirality and the same number $N_s$ of spinors with $-$ chirality \cite{footnoteP}. We observe that only  minuscule  representations -- spinors and elementary
fermions survive in our limit. 
The high rank antisymmetric states ($V_r$ with $r$ of the order of $N$) disappear from the spectrum, because they're too heavy. A lower rank 
 tensors   are decoupled into $r$ elementary fermions. 
 
  Elementary fermions interact to tehmselves via quantum fluctuations of $\Delta $ represented by the GN-model, but in the adiabatic limit $\Delta$ is frozen:   vector particles interact with kinks, but not to themselves.
We clearly see it from \eq{82}:  $K_{\vv\vv}$ and all  $K_{rr'}$ with light  $r/h^\vee,r'/h^\vee\to 0$ vanish, and 
$K_{rs}\to rK_{\vv s}$ as follows from (\ref{84},\ref{88}).    We see it also from  the mass spectrum formula \eq{191}\!.
At $r/h^\vee\to 0$ and $m_{v}$
kept fixed, the mass spectrum stops being concave as $m_r\to  r m_{v} $. It is consistent
 with the fusion rule \eq{91}. As the pole of the
vector-vector scattering  $\theta=\vartheta_1\to i\pi$  disappears from the physical
sheet, the fusion rule prompts  $s_{\vv\vv}(\theta)\to1$.   At the same time the pole  of the fermion-spinor scattering $\mathfrak{X}_{\vv\pm}$
in \eq{77} equall to $\theta=\vartheta_1/2\to i\pi/2$ remains on the physical sheet: vectors keep forming bound states with
kinks. We have seen  in the P\"oschl
-Teller model of Sec.~\ref{PT} : a pole in
the scattering amplitude  \eq{381} corresponds to the Majorana zero mode, a bound state
of the fermion and the kink. 
 Comparing with the classical case, we identify $m_{\vv}$ with $\Delta_0 $ in \eq{811}
\begin{align}\Delta_0=m_{\vv}=\underset{\VM{h^\vee}\to\infty}\lim(2\pi/h^\vee )m_{s}\,\la{94}
\end{align}
\VM{We recall that the spinor mass $m_s$ grows linearly with $N$, so the limit is in \eq{94} is finite. }
     
The limit of the vector-spinor scattering  obtained from the explicit quantum formulas (\ref{79},\ref{771}) is in agreement with the P\"oschl
-Teller amplitude \eq{171}
\begin{align}
     &s_{\vv\pm}\to \pm A(\theta)=\pm\tanh\left(\frac\theta 2+\frac{i\pi}4\right)\,\la{96}
\end{align}
We remark that the limiting values of the building blocks of \(s_{v\pm}\) are  $X_{\vv\pm}=\pm\mathfrak{S}_{\vv s}^{-2}=\mp A^2$.
 To compare the  P\"oschl
-Teller scattering with   quantum formulas we recall
the representation of projectors in \eq{701} in terms of $O(2N)$ gamma-matrices:
  $\left(\mathscr{P}_{\pm}\right)_{\mu\nu}=\tfrac 1{2(N-2)}(\delta_{\mu\nu}\mp\Sigma_{\mu\nu})\Gamma_\pm$,
where    $\Sigma_{\mu\nu}=\tfrac 12[\gamma^\mu,\gamma^\nu]$ is the spin operator.
At  $N=2$, that corresponds to the P\"oschl--Teller, $\mathscr{P}_{\pm}$ vanishes
 as in this case $1\mp\Sigma=\Gamma_\mp$. 
Then  $\mathscr{P}_{\vv\mp}=1$ and the S-matrix given by \eq{701} is reduced
to $\pm\, A(p)$ in agreement with  \eq{381}.

Hence,
\begin{align}
\lim_{h^\vee\to\infty}K_{\vv s}= 
 -\frac{1}{2\pi\cosh\theta}\,.
\label{98}
\end{align}

Now we turn to the limit of the spinor-spinor scattering.    
We see that leading large $N$ order of the spinor-spinor scattering is  no longer meromorphic and, as such, does not represent particle scattering any longer. It has a branch cut along the real axis. Defining
 \begin{align}
k_{ss} :=\underset{h^\vee\to\infty} \lim \tfrac 1{\ h^\vee}\tilde K_{ss}\,, \la{941}
\end{align}
and  taking the limit of   \eq{73} we obtain
\begin{align}
k_{ss}(\theta)=-\tfrac
1{(2\pi)^2}\int_{0}^\infty  
 \frac{ 1}x{\tanh \frac{\pi x}{2}}\,\cos (x\theta)dx=-\frac {1}{  2\pi^2}\log \left|\coth\frac \theta 2\right|\,.\la{931}
\end{align}
In the next section, we show how this formula comes from fusion relations. 

The regular part of the  scattering phase that corresponds to  \eq{931} is  expressed in terms of   the Euler dilogarithm ${\rm Li}_2(z)=\sum_{n=\VM{1}}^\infty\frac{z^n}{n^2}$
\begin{align}
  \lim_{h^\vee\to\infty}\frac \pi {\ h^\vee}\left[\Theta_{ss}(\theta)-\Theta_{ss}(\infty)\right]\to  \VM{ {\rm Li_2}\left(e^{-\theta}\right)-{\rm Li_2}\left(-e^{-\theta}\right),\quad \theta>0\,}.
\label{}
\end{align}
We repeat that the limiting value of the spinor-spinor scattering cannot be associated with a scattering problem and does not seem to have an obvious analog in the classical soliton theory. In the quantum theory, kinks are true particles \footnote{\VM{Euler dilogarithm often appears as a semiclassical limit of quantum S-matrices, such as the kink–kink S-matrix of the sine-Gordon model \cite{Faddeev}. The kink–kink $D_N$ S-matrices discussed here are different from the quantum dilogarithms of Ref.~\cite{Faddeev}, despite that both have a similar semiclassical limit expressed in terms of Euler dilogarithm.}
}

We observe a similar pattern if
the ground state is made out of heavy antisymmetric tensors with rank $r^\ast:=\tfrac 12(h^\vee-2r)$. Then, assuming $r,r'\sim 1$ the formula \eq{88} promts
 \begin{align}
\lim_{h^\vee\to\infty} K_{r^\ast r'}= k_{\vv s}, \quad  
\underset{h^\vee\to\infty} \lim \tfrac 1{\ h^\vee} \tilde K_{r^\ast {r'}^\ast}= 2 k_{ss}\,. \label{981}
\end{align}
Once again, this illustrates that the semiclassical limit obscures many subtle features of the quantum theory.

\subsection{Semiclassical limit of fusion relation}
It is instructive to obtain the limiting values of \(K_{ss}\)  (\ref{931}) without using the explicit formulas \eq{46}.  follows  from the fusion formulas  Sec.~(\ref{5.4}).

Take the fusion relation in the form of \eq{901} and expand its LHS in $2i\pi/h^\vee$  and use the property $k_{\vv s}(\theta)=-k_{\vv s}(\theta+i\pi)$.
Then \eq{901} prompts   the relation
\begin{align}2\pi\tfrac {d}{d\theta} k_{ss}(\theta)=-ik_{\vv s}\left(\theta+\frac{i\pi}2\right)=
\frac{1}{\pi\sinh\theta}\,.\la{100}
\end{align}
We know its RHS  from the solution of the P\"oschl-Teller problem. Integration prompts the formula \eq{931}. We observe that the kernels  in the pairs of equations (\ref{106},\ref{110}) for vector particles and   for  kinks  are related, \la{114}
so the pairs of equations are not independent.

\section{Semiclassical limit of the thermodynamic Bethe equations \la{CB}}
Given the scattering, we are equipped to address the spectrum. 
The spectrum is determined by the thermodynamic Bethe equations quoted at the end of Sec. \ref{FG} with $K_{ss}$ and $K_{vs}$ given by \VM{(\ref{K},\ref{771})}. 
We recall that these equations describe the spectrum of a state that consists of equal number  of spinors of opposite chirality occupying a single interval. Now we study the semiclassical limit of these equations. 
   
We begin with a general remark on the Bethe equation \eq{31}, which we repeat here
\begin{align}
\frac{d}{d\theta}\begin{pmatrix}P_{s} \\
E_{s} \\
\end{pmatrix} =m_{s}\begin{pmatrix}\cosh\theta \\
\sinh\theta \\
\end{pmatrix}+ \int_{-B}^B K_{ss}(\theta-\theta')\,d
\begin{pmatrix}P_{s}  \\
E_{s}   \\
\end{pmatrix} \,. \label{97}
\end{align}
The equation holds for an arbitrary real $\theta$. However, the meaning of the equation  depends on the position of  $\theta$, as was emphasized  in Ref.~\cite{Yang} (where thermodynamic
Bethe equation was introduced).   If \(\theta\) is chosen within the interval $(-B,B)$, then the term in the LHS   is merely the density of states $(2\pi/L)\rho_s$, see Sec.~\ref{FG}. In this case, we write \eq{97} as
 \begin{align}
|\theta|<B:\quad-\int_{-B}^B \tilde K^{\rm }_{ss}(\theta-\theta')\,
d\begin{pmatrix}P_{s}  \\
E_{s}  
\end{pmatrix}= m_{s}\begin{pmatrix}\cosh\theta \\
\sinh\theta \\
\end{pmatrix}\,,\la{9811}
\end{align}where \(\tilde K_{ss}\) is defined in \eq{K} and explicitly given by \eq{73}. This equation alone determines the density of the occupied states. However, if $\theta$ is outside of the interval, where there is no particles,  the LHS   describes a density of  "holes", an unoccupied part of the spectrum suitable for excitations. Once the density of occupied states is found by solving the Bethe equations \eq{9811}within the interval, the density of "holes" is determined by the  RHS of  \eq{97} evaluated outside of the interval.  

In a regular situation, the particle branch and "hole" branch are smoothly connected, forming a smooth $P_s(\theta)$.
The semiclassical limit is singular, and this is the essence of the Peierls phenomenon. We see  that $\tilde K_{ss}$  and the mass of a spinor particle $m_s$  are of the order of $N$ (see (\ref{94},\ref{102})), and balance each other in \eq{9811}. However, if $|\theta|>B$, there is an imbalance, since the LHS in \eq{97} is of the order 1 \VM{while the RHS is of the order $N$}. 
\VM{In the large $N$ limit, there is no longer a single continuous function describing the solution to the Bethe equations both outside and inside the interval $[-B,B]$.}
The large $N$ limit truncates the Bethe equations by forcing the spinor branch
to be fully occupied, in complete agreement with the Peierls phenomenon:  the Fermi momentum is the endpoint of the spectrum. 

Dividing  both sides of \eq{9811} by \(h^\vee\), taking the limit  and using  (\ref{94},\ref{941}) we obtain  the   semiclassical Bethe  equations
\be
|\theta|<B:\quad
-\int_{-B}^B k_{ss}(\theta-\theta')d\begin{pmatrix}P_{s}  \\
E_{s}  
\end{pmatrix} =  \frac{\Delta_0}{2\pi}\begin{pmatrix}\cosh\theta \\
\sinh\theta \\
\end{pmatrix},\quad k_{ss}(\theta)=-\frac {1}{  2\pi^2}\log \left|\coth\frac \theta 2\right|\,.\label{106}
\ee

This agrees with the Peierls phenomenon: the spectrum ends at the Fermi momentum.  

Next, we turn to the Bethe equations for the vector branch. The Bethe  equations are specifications of (\ref{184},\ref{205}) for \(q=\vv,\, q'=\pm\). 
In this case, all the terms are of the same order.  Taking the   limit, we obtain 
 \begin{align}
\begin{aligned}
\frac{d}{d\theta} \begin{pmatrix}P_{\vv}  \\
E_{\vv}  
\end{pmatrix}= \int_{-B}^B k_{\vv s}(\theta-\theta')\,
d\begin{pmatrix}P_{s}  \\
E_{s}  
\end{pmatrix}\, &+ \Delta_0\begin{pmatrix}\cosh\theta \\
\sinh\theta \\
\end{pmatrix}\,, \label{110}
\end{aligned}
\end{align}
where \(k_{\vv s}\) follows from \eq{98}
\begin{align}k_{\vv s}(\theta):=2\underset{h^\vee\to\infty}\lim K_{\vv s}(\theta)= -\frac{1}{\pi\cosh\theta}\,.\la{113}
\end{align}

These equations describe an occupied part of the spectrum. It should be clear that the two branches are interdependent. Once the occupied part of the spectrum is known (by solving Eq.\eq{106}), then the evaluation of the RHS of \eq{110} yields the spectral branch of the vector particles. It is noteworthy that  $k_{ss}$ and $k_{\vv s}$ are
interconnected singular kernels (\ref{98},\ref{102}). 
The remaining part of the paper dwells on this property.

Let us differentiate \eq{106} and 
 use the fusion relation \eq{100}. We obtain 
 \begin{align}
 |\theta|<B:\quad -\int_{-B}^B k_{\vv s}\left(\frac{i\pi}2+\theta-\theta'\right)\,
d\begin{pmatrix}P_{s}  \\
E_{s}  
\end{pmatrix} =i\Delta_0\sinh\theta\,
\label{101}
\end{align}
and compare \eq{101} with \eq{110}. The arguments below equally apply to momentum and energy. We write the momentum equation explicitly:  
\begin{align}
&|\theta|<B:\quad \quad\qquad 0=\frac 1\pi\int_{-B}^B \frac{dP_s}{\sinh(\theta-\theta')} +\Delta_0\sinh\theta\,,\la{102}\\
 &|\theta|>B:\quad\tfrac{d}{d\theta}\ P_{\vv}(\theta)= \frac 1\pi\int_{-B}^B \frac{dP_s}{\cosh(\theta-\theta')}\, + \Delta_0\cosh\theta\,.\la{103}
\end{align}

The  equation \eq{102} is a standard singular integral equation that defines
$2\pi i$-periodic analytic function \(f(z)\) in the complex plane, cut along the interval
$[-B, B]$.  The boundary value of this function on the interval is
\begin{align}
|\theta|<B:\quad f(\theta)d\theta =d(iP_s(\theta)+\Delta_0\cosh\theta)\,.
\end{align}
Define an analytic function on the physical sheet $0<{\rm Im}\ z<\pi$ 
\begin{align}F(\VM{z})=\frac 1\pi\int_{-B}^B \frac{\IM f(\theta)d\theta}{\sinh(\VM{z}-\theta)}
+\Delta_0\sinh \VM{z}\,.\end{align}  Since, \(\Delta_0\cosh\theta\)  is also a boundary value of  an  analytic function \(\Delta_0\cosh z\), \(dP_s(\theta)\) can also be analytically extended. We denote it by \(dP_s(\VM{z})\).
Therefore, \(F(\VM{z})d\VM{z}=idP_s(\VM{z})\). Now we notice that \eq{103} implies the relation between momenta differentials of elementary fermions.   A similar relation holds for the energy:  energy-momentum of the elementary fermions band treated as a function of rapidity is obtained by the analytic continuation of the energy-momentum of the central band from the interval of the real rapidity $\IM\theta=0,\ |\theta|<B$.
The central band formed by the hybridized zero modes is the band of the elementary fermion. It is a  different {\it real sections} of a single spectral curve sliced by the line $\IM\theta=0$ and $\IM\theta=\pi/2$ respectively.  Naturally, the spectral density of each branch is given by the same Abelian differential, which we now compute. This miracle, of course, does not occur at finite $N$.

\subsection{Spectral curve \la{7.3}}

The semiclassical Bethe equations are singular integral equations with Cauchy-type kernels. These equations can be formulated as a Riemann-Hilbert problem and, in the case of a single interval, admit explicit solutions.
Let us take the equation for the energy~(\ref{106}) and integrate it by parts. In this way, we obtain a single integral equation
\begin{align}
-\VM{\text{p.v.}}\int\limits_{-B}^{B} \frac{g(\theta')}{\sinh(\theta - \theta')} \frac{d\theta'}{\pi}
=  \Delta_0 \sinh \theta\,,
\label{107}
\end{align}
where $g(\theta) d\theta$ is either $dP_s$ or $E_s(\theta)d\theta$ in the  $P>0,\ E<0$ quadrant of the spectrum. Other quadrants are obtained by reflections. 
Solutions inside the interval, with at most integrable singularities at the endpoints, are given by
\begin{align}
    g(\theta) = \frac{C}{\sqrt{\sinh^2 B - \sinh^2\theta}}
    -  \Delta_0\sqrt{\sinh^2 B - \sinh^2\theta},
    \label{115}
\end{align}
with a real constant $C$ yet to be determined.

We begin with the energy $E_s$.  
Since the energy must remain finite, we set $C=0$.  
Furthermore, because the kink band  (the central band ) is formed by hybridized fermionic zero modes localized on kinks, the spectrum is symmetric with respect to  $E=0$. That fixes the additive constant.  
Hence, the kink's energy vs rapidity reads 
\begin{align}
E^2_s(\theta)=\Delta_0^2(\sinh^2 B - \sinh^2\theta)=E_-^2 - p^2\,.
\la{1071}
\end{align}
where we denoted
\begin{align}
  E_+ \pm E_- = \Delta_0\, e^{\pm B}\,.
\end{align}
We see that $\pm E_-$ are the endpoints of the central band and therefore the endpoints of the gaps.
We emphasize that the rapidity $\theta$  or  $p$ in \eq{1071}, do no longer corresponds to asymptotic states, as it happens before the large $N$ limit is taken when all bands extend to infinite energy. Later, we will see that $\pm E_+$ is another endpoint of the gaps.

We now turn to the momentum.  
It features a van Hove square-root singularity at the endpoints, and in this case $C\neq0$: 
\begin{align}
    dP_s = -\left(C/E_s (\theta)- E_s(\theta\right))d\theta.
    \label{118}
\end{align}

Using~\eqref{1071}, we express the mass-shell measure $d\theta = dp^0/p$ (see~\eqref{22}) in terms of $E_s$:
\begin{align}
  d\theta =- \frac{E_s\, dE_s}{y(E_s)}\,,
\end{align}
where
\begin{align}
  y ^2=(E_-^2 - E_s^2)(E_+^2 - E_s^2)\,.
\label{120}
\end{align}
\VM{Combining} (\ref{118}-\ref{120}) we see that  the spectral density is given by  an Abelian differential of the second kind on the elliptic curve defined by~\eqref{120},  as expected from the algebro-geometric approach (see, Sec.~\ref{1.2.2})
\begin{align}
  dP_s = \,\frac{C - E^2}{y(E)}\, dE\,.
\label{122}
\end{align}
 
 The most lower band (and the most upper band)  in Fig.~\ref{spectrum} correspond to elementary fermions.
We analytically continue the central band energy \eqref{1071} to the line \(\VM{z}=\theta + \frac{i\pi}{2}\), obtaining
\begin{align}
  E_{\vv}^2
  = {\Delta_0^2\cosh^2\theta + E_-^2}
  = {E_+^2 + \Delta_0^2\sinh^2\theta}\,.
\end{align}
Equivalently,
\begin{align}
  E_{\vv}^2 = E_+^2 + p^2\,.
\label{121}
\end{align}
Hence, $\pm E_+$ is the endpoint of the elementary fermion bands and the endpoints of the gaps. 
No wonder that the analytical form  of the  differential of  momentum   for the elementary fermion branch
\be
dP_{\vv} = \,\frac{C - E^2}{y(E)}\, dE
\ee
is of no different from that of the central band. Different branches of the spectrum are real sections of a single complex curve.  \VM{This result illustrates a general feature of soliton theory: the multiple branches of the spectrum of finite-gap solutions are branches of the real section of a single algebraic curve. Our particular example indicates the quantum origin and the quantum prototype of this phenomenon. Namely, it arises from the relation between the semiclassical limits of the spinor–spinor and vector–spinor scattering phases represented by Eq.~\eqref{100}. The latter, in turn, originates from the fusion formula \eqref{901} solely encoded in the Dynkin scheme $D_N$.}

The remaining task is to express the parameters $E_\pm$ and $C$ in terms of the number of particles \(N_s\).  
Similar to the algebro-geometric approach of the Sec.~\ref{1.2.2}, these parameters follow from  the conditions  given by  Eq.~\eqref{18}  
\begin{align}
\int_{-E_-}^{E_-} dP_s =(2 \pi/L) N_s,\quad \int_{E_-}^{E_+} dP_s=0 \,.\la{119}
\end{align}
Together they yield the expressions quoted in Eq.~\eqref{2211}.
\footnote{In Ref. \cite{MWZ}, we did not explore the second condition of \eq{119}.  The nessesary information was extracted by evaluating the first equation of~\eqref{106}
at $\theta = 0$ that brings the same result.}
%
%

\section{Summary}

Let us summarize the main results of this work.

We studied the large-rank limit of an integrable quantum field theory with Lie-group symmetry in the presence of a specific chemical potential. Our analysis focused on the Lie group \(O(2N)\).  The chemical potential was chosen so that the ground state is the thermodynamic state composed of spinor particles. The  scattering  of such theory,
as well as everything else,   is defined entirely by the Dynkin scheme of
type \(D_N\).

Our primary goal was to determine the limiting form of the particle spectrum as the rank of the Lie algebra increases.  We demonstrated that the spectrum coincides with the finite-gap potential of the Dirac equation previously obtained by algebro-geometric methods in soliton theory. In this context, the Dirac operator acts as the Lax (Zakharov–Shabat) operator generating classical soliton equations, such as the modified Korteweg–de Vries (mKdV) equation and the increasing Coxeter dual number serves as a semiclassical parameter.

A central observation of this work is the mechanism by which meromorphic differentials, fundamental objects in the algebro-geometric approach to soliton theory, emerge in the large-rank limit. Before the limit is taken, the differentials of momentum and energy are not meromorphic, as they should not be in a regular quantum theory. We illustrated this phenomenon using the simplest nontrivial finite-gap potential: the traveling-wave solution of the defocusing mKdV equation, corresponding to a two-gap spectrum and an associated elliptic curve. We do not anticipate conceptual obstacles to extending our approach to more complicated finite-gap potentials and also to different large rank Dynkin scheme.

At the same time, the methods of quantum integrable models employ several structures whose classical counterparts in soliton theory have not yet been identified. In our framework, the spectrum arises from the large-rank limit of the thermodynamic Bethe equations, which take a form quite different from those used in classical finite-gap theory. Before the limit, the Bethe equations describe a smooth quantum spectrum. However, in the large-rank limit, these equations degenerate into singular integral equations, whose solutions are  Abelian meromorphic differentials. We do not know of a classical analogue of the singular Bethe equations within soliton theory.

Another conceptual issue concerns the role of spinor particles in the classical theory of solitons. The Bethe equations are built from the scattering amplitudes of spinor particles. In the classical limit, the objects that correspond most closely to spinors are kinks—the half-soliton solutions of the mKdV equation. Yet kinks are not particles in the classical sense,  and the limiting form of the spinor scattering amplitude does not have an obvious interpretation as kink–kink scattering. Nevertheless, this amplitude enters the Bethe equations and ultimately determines the finite-gap spectrum. Its precise meaning within the classical soliton theory remains unclear. The same can be said about the role of the fusion relations in the classical setting.

These features, like nearly all other structural elements discussed in this paper, originate from the Lie algebra of the underlying Lie-group symmetry of the quantum theory. It is therefore natural to expect that in the classical limit, infinite-rank Lie algebras should also emerge as the relevant organizing structures, even if their role has not yet been identified. 

\vspace{-0.3cm}

\section*{Acknowledgment}
We would like to thank Alexander Its for his insightful discussions. The work of K.Z. was supported by a VR grant 2021-04578.  The work of V.M. was supported by VR grant 2023-04726 and the Roland Gustafsson Foundation. The work was supported by the NSF under Grant NSF
DMR-1949963.  P. W. gratefully acknowledges the hospitality at Nordita, where this work was conducted. The work was reported at the 2024 BIMSA  workshop Integrable Systems and Algebraic Geometry, dedicated to the memory of Igor Krichever. 


\begin{thebibliography}{99}
\bibitem{FootnoteSimple} The Lie group symmetry determines solutions of integrable models even if the algebra is semi-simple, see e.g., \cite{ORW}.
\bibitem{FootnoteM} The minuscule representation is the irreducible representation such that the Weyl group acts transitively on the weights.
\bibitem{footnoteP} Quantum states that correspond to  other periodic solutions, such as kink breathers on top of the traveling wave \cite{Pelinovsky}, are readily available in the quantum theory, but are
 not discussed here.
\end{thebibliography}
\end{document}